\documentclass[twocolumn]{aastex631}

\graphicspath{{./}{figures/}}
\usepackage{amsmath}

\usepackage{CJK}

\newcommand\sref[1]{\hyperref[#1]{\S~\ref*{#1}}}
\newcommand\fref[1]{\hyperref[#1]{Fig.~\ref*{#1}}}
\newcommand\Eqref[1]{Eq.~(\hyperref[#1]{\ref*{#1}})}
\newcommand\eeqref[1]{Eq.~\hyperref[#1]{\ref*{#1}}}
\newcommand\tref[1]{\hyperref[#1]{Table~\ref*{#1}}}
\newcommand\aref[1]{\hyperref[#1]{Appendix~\ref*{#1}}}

\shorttitle{Bridging Scales: Galactic nucleus to cosmic environment}
\shortauthors{Su et al.}

\begin{document}
\begin{CJK}{UTF8}{mj}
\title{Bridging scales: How much do supermassive black holes grow in the suppressed Bondi regime?}

\author[0000-0003-1598-0083]{Kung-Yi Su}
\affiliation{Department of Physics \& Astronomy and Center for Interdisciplinary Exploration and Research in Astrophysics(CIERA), Northwestern University, 1800 Sherman Ave, Evanston, IL 60201, USA}
\affiliation{Black Hole Initiative at Harvard University, 20 Garden Street, Cambridge, MA 02138, USA}
\affiliation{Center for Astrophysics $\vert$ Harvard \& Smithsonian, 60 Garden Street, Cambridge, MA 02138, USA}
\email{kungyisu@gmail.com}

\author[0000-0001-5287-0452]{Angelo Ricarte}
\affiliation{Black Hole Initiative at Harvard University, 20 Garden Street, Cambridge, MA 02138, USA}
\affiliation{Center for Astrophysics $\vert$ Harvard \& Smithsonian, 60 Garden Street, Cambridge, MA 02138, USA}
\email{angelo.ricarte@cfa.harvard.edu}

\author[0000-0002-5554-8896]{Priyamvada Natarajan}
\affiliation{Black Hole Initiative at Harvard University, 20 Garden Street, Cambridge, MA 02138, USA}
\affiliation{Department of Astronomy, Yale University, Kline Tower, 266 Whitney Avenue, New Haven, CT 06511, USA}
\affiliation{Department of Physics, Yale University, P.O. Box 208121, New Haven, CT 06520, USA}
\email{priyamvada.natarajan@yale.edu}

\author[0000-0002-1996-0445]{Antonio J. Porras-Valverde}
\affiliation{Department of Astronomy, Yale University, Kline Tower, 266 Whitney Avenue, New Haven, CT 06511, USA}
\email{antonio.porras@yale.edu}

\author[0000-0002-2858-9481]{Hyerin Cho (조혜린)}
\affiliation{Center for Astrophysics $\vert$ Harvard \& Smithsonian, 60 Garden Street, Cambridge, MA 02138, USA}
\affiliation{Black Hole Initiative at Harvard University, 20 Garden Street, Cambridge, MA 02138, USA}

\author[0000-0002-1919-2730]{Ramesh Narayan}
\affiliation{Center for Astrophysics $\vert$ Harvard \& Smithsonian, 60 Garden Street, Cambridge, MA 02138, USA}
\affiliation{Black Hole Initiative at Harvard University, 20 Garden Street, Cambridge, MA 02138, USA}

\author[0000-0002-4900-6628]{Claude-Andr\'e Faucher-Gigu\`ere }
\affiliation{Department of Physics \& Astronomy and Center for Interdisciplinary Exploration and Research in Astrophysics(CIERA), Northwestern University, 1800 Sherman Ave, Evanston, IL 60201, USA}

\author[0000-0003-3729-1684]{Philip F. Hopkins}
\affiliation{TAPIR, Mailcode 350-17, California Institute of Technology, Pasadena, CA 91125, USA}

\author[0000-0002-0393-7734]{Ben S. Prather} 
\affiliation{Black Hole Initiative at Harvard University, 20 Garden Street, Cambridge, MA 02138, USA}

\begin{abstract}
\label{abstract}
The co-evolution of supermassive black holes (SMBHs) and their host galaxies remains one of the central open questions in cosmology, rooted in the coupling between accretion, feedback, and the multi-scale physics that links the event horizon to the circumgalactic medium. Here we bridge these scales by embedding a first-principles, GRMHD-informed prescription for black hole accretion and feedback--derived from multi-zone simulations that self-consistently connect inflows and outflows from the horizon to the Bondi radius--within cosmological magnetohydrodynamic zoom-in simulations of $\sim10^{14}\,M_\odot$ halos. These GRMHD results predict a ``suppressed Bondi'' regime in which magnetic stresses and relativistic winds strongly reduce effective accretion rates in a spin-dependent manner. We find that black holes cannot grow efficiently by accretion until they exceed $\sim10^{7}\,M_\odot$, regardless of the feedback strength. Beyond this threshold, systems bifurcate: low-spin ($\eta\!\sim\!0.02$) black holes continue to accrete without quenching star formation, while high-spin ($\eta\!\gtrsim\!0.3$) black holes quench effectively but become starved of further growth. Early, massive seeding partially alleviates this tension through merger-driven assembly, yet an additional cold or super-Eddington accretion mode appears essential to reproduce the observed SMBH population and the empirical black hole--galaxy scaling relations. Our results demonstrate that GRMHD-informed feedback models can account for the maintenance-mode behavior of low-luminosity AGN like M87*, but cannot by themselves explain the full buildup of SMBH mass across cosmic time. A unified, multi-regime framework is required to capture the evolving interplay between spin-dependent feedback, cold inflows, and mergers in driving co-evolution.
\end{abstract}
\keywords{Accretion (14), Active galactic nuclei (16), Bondi accretion (174), Schwarzschild black holes (1433), Supermassive black holes (1663), Magnetohydrodynamical simulations (1966)}

\section{Introduction}
\label{S:intro}
The physical mechanisms that channel gas from galactic scales into the deep potential wells of supermassive black holes (SMBHs), and in turn allow their energetic output to regulate galaxies across cosmic time, remain elusive. Accretion onto black holes -- observed as active galactic nuclei (AGN) -- is widely recognized as a dominant agent of feedback, capable of halting star formation and sustaining the long-lived quiescent phases of massive galaxies that populate the ``red and dead’’ sequence. Yet, the exact pathways by which AGN couple their immense energy and momentum to their environments remain one of the central challenges in our theoretical understanding of galaxy evolution, particularly in the most massive halos with $M \gtrsim 10^{12}\,M_{\odot}$, where traditional feedback models struggle to reproduce the observed suppression of star formation \citep[e.g.,][]{Bell+2003,Kauffmann+2003,Madgwick+2003,Baldry2004,keres+2005,Blanton+2005,Dekel+2006,keres+2009,Pozzetti+2010,Wetzel+2012,Feldmann+2015,Voit+2015}.
Observations across the electromagnetic spectrum, and especially in X-rays, show that massive ellipticals and cluster cores contain significant reservoirs of hot gas \citep[e.g.,][]{Fabian+1994,Peterson+2006,McDonald+2011,Werner+2013,Stern+2019,Mercedes+2023}. Left unchecked, this gas is likely to cool and condense into the galactic center, likely triggering vigorous star formation. However, such starbursts are conspicuously absent \citep{Tamura+2001,ODea+2008,Rafferty+2008} (except in a few cases like the Phoenix cluster \citealt{McDonald+2012}), producing what is referred to as the long-standing “cooling flow problem.” Evidently, star formation is being effectively suppressed in the nuclei of massive galaxies. Alternative quenching methods such as stellar feedback or cosmic ray pressure appear to be insufficient on their own \citep{Su+2019}, pointing to AGN feedback as the primary causal mechanism behind the observed regulation.

Progress in elucidating the workings of AGN feedback has been hampered by the lack of a first-principles treatment of the underlying microphysics, the scarcity of direct observational constraints, and the enormous computational gulf between the physical scales involved. Consequently, cosmological simulations of galaxy formation continue to rely on sub-grid prescriptions for black hole accretion and feedback--phenomenological models that encapsulate unresolved processes through empirical scaling relations \citep[e.g.,][]{Li2014,Fiacconi2018,AnglesAlcazar+Zoom_2021,Talbot2021,Weinberger2023}, and variants of such recipes are now standard across the several simulation suites \citep[e.g.,][]{Sijacki2015,Rosas-Guevara2016,AngleAlcazar+2017,Weinberger2018,Ricarte2019,Ni2022,Wellons2023,2024ApJ...973..149B}.

At the opposite extreme, general relativistic magnetohydrodynamic (GRMHD) calculations that can follow the accretion and feedback down to the Event Horizon, find that these processes naturally generate relativistic jets and winds \citep[e.g.,][]{Gammie2003,Tchekhovskoy2011,Porth:2019,Chatterjee:2023}. Yet such simulations remain highly idealized since embedding them within a cosmological environment is prohibitively computationally expensive. Among these efforts, a new generation of multi-zone GRMHD frameworks has emerged, capable of coupling inflows and outflows across the vast dynamical range that bridges the event horizon and galactic scales. By iteratively evolving each zone on its characteristic timescale, these models achieve, for the first time, a self-consistent, dynamically converged description of accretion and feedback spanning more than seven orders of magnitude in scale. This breakthrough now enables the direct infusion of first-principles physics from horizon-resolved simulations into cosmological contexts, providing a long-sought bridge between black hole microphysics and galaxy evolution \citep[e.g.,][]{Ressler2020,Lalakos2022,Guo2023,Guo+2024,Guo+2025,Hopkinsforgedinfire2023,Hopkins+2025timedialation,Cho+2023,Cho+2024,Kaaz2024}. 

Amongst these efforts, \citet{Cho+2023,Cho+2024} proposed a novel multi-zone framework that enables bidirectional coupling across the enormous dynamic range relevant to accretion and AGN feedback for the first time.
Deploying this method, smaller scales are repeatedly frozen and only intermittently evolved, a strategy that circumvents the otherwise prohibitive horizon-scale timestep constraints. Each scale evolves according to its own time scale, allowing enough time for each scale to respond to any information from smaller or larger scales. By repeatedly sweeping through the hierarchy of zones--cycling upward and downward across the radial annular structure hundreds of times--the method achieves, for the first time, a numerically converged dynamical steady state for both inflows and outflows spanning more than seven orders of magnitude in scale. In \citet{Cho+2025}, this framework was further generalized to include cases with non-zero black hole spin, wherein relativistic jets dominate the feedback power. The feedback efficiencies $\eta$ for these models were found to be insensitive to the Bondi radius but dependent on the black hole spin $a_*$, where $\eta\sim 0.3$ for $a_*=0.9$ and $\eta\sim 0.02$ for $a_*=0$. Together, these studies provide a first--principles-informed AGN feedback model, with the feedback efficiency measured at the Bondi radius, making it directly applicable to galaxy-scale simulations.

The first full cycle linking galaxy simulations down to horizon-resolving GRMHD simulations and back was completed together in \citet{Cho+2024} and \citet{Su+2025Bridging}. In \citet{Cho+2024}, the multi-zone's large scales were initialized from an M87-like isolated galaxy simulation. Once the multi-zone GRMHD run converged, yielding inflow and outflow profiles from the Bondi radius down to the horizon, the resulting suppressed accretion and feedback efficiency were subsequently fed back into the same galaxy simulation in \citet{Su+2025Bridging}.
A key conclusion for galaxy evolution was that if the SMBH has already assembled its $z=0$ mass, then even the minimal feedback efficiency predicted by the non-spinning ($a_\ast=0$) case is sufficient to maintain black hole mass stability over gigayear timescales. Moreover, for feedback efficiencies $\eta\equiv \dot{E}_{\rm wind}/\dot{M}_{\rm BH} c^2 \gtrsim 0.15$--requiring moderate black hole spin--the accretion rate can be regulated to those of M87* and Sgr A* inferred from the Event Horizon Telescope \citep{EHTm87_8_2021, EHTMW_5_2022} observations. The remaining open question is whether a black hole can simultaneously grow to its present-day mass and suppress star formation if the multi-zone GRMHD-informed AGN feedback from \cite{Cho+2025} are assumed to operate.  In particular, given that the multi-zone GRMHD simulations best match low-Eddington systems, it is crucial to explore how broadly this feedback mode applies.

To address this question, in this paper, we implement the multi-zone-GRMHD-informed AGN feedback models within a cosmological zoom-in simulation of a halo with $z=0$ mass $\sim10^{14} M_\odot$. The setup includes suppression of BH accretion with respect to the Bondi accretion rate, with feedback efficiencies calibrated to GRMHD results obtained for different black hole spins. We survey models informed by various spin states and seeding prescriptions, testing the conditions under which a black hole can simultaneously grow in mass and quench star formation in its massive host galaxy. Furthermore, we examine when this mode of accretion can dominate the overall growth of the black hole.

\section{Methodology}
\label{S:methods}

We perform cosmological zoom-in simulations of a galaxy in a $z=0$ halo of $10^{14}\,M_\odot$ with the GIZMO code\footnote{A public version of GIZMO is available at \href{http://www.tapir.caltech.edu/~phopkins/Site/GIZMO.html}{\textit{http://www.tapir.caltech.edu/$\sim$phopkins/Site/GIZMO.html}}}
 \citep{Hopkins2015mfm}. We run GIZMO in mesh-less finite-mass (MFM) mode, a Lagrangian mesh-free Godunov hydrodynamics scheme that combines the strengths of grid-based and smoothed-particle hydrodynamics (SPH). Numerical algorithms and code tests, including hydrodynamics, self-gravity, and magnetohydrodynamics (MHD), are described in \citet{Hopkins2015mfm,Hopkins+2015mhd_divergence,Hopkins+2016mhd}.
 
All simulations use the FIRE-2 framework for gas cooling, the interstellar medium (ISM), star formation, and stellar feedback (see \aref{a:sfb}), with finest gas mass resolution $\sim 4\times 10^{5}\,M_{\odot}/h$ and jet mass resolution $\sim 5\times 10^{4}\,M_{\odot}/h$. The methodology and its numerical tests are described in detail in \citet{Hopkins+FIRE2018}; we summarize only key elements here. FIRE-2 follows radiative cooling from $10$ to $10^{10}\,\mathrm{K}$, and stellar feedback from supernovae, stellar winds, and radiation. Star formation uses a sink-particle prescription restricted to gas that is dense, molecular, self-shielding, and locally self-gravitating.

Black hole accretion is modeled based on the Bondi prescription,
\begin{align}\label{eq:bondi}
\dot{M}_{\rm Bondi} = \frac{4\pi \rho G^2 M_{\rm BH}^2}{(c_s^2 + \Delta v^2)^{3/2}},
\end{align}
where $\rho$ is the local gas density, $c_s$ is the mean sound speed, and $\Delta v$ is the mean relative velocity near the BH. These quantities are evaluated by averaging within a kernel enclosing 96 effective neighbors, subject to a minimum radius of 50 comoving pc/$h$ (see \aref{a:acc}).

The actual BH accretion rate is scaled from the Bondi rate by a prefactor $f_{\rm mass, acc}$, following the multi-zone GRMHD-informed model of \citet{Cho+2024,Cho+2025}: 
\begin{align}\label{eq:kaappa_cho}
\dot{M}_{\rm BH}&=f_{\rm mass, acc} \dot{M}_{\rm Bondi}\notag\\ f_{\rm mass, acc}&=\kappa_{\rm cho}\equiv \left(\frac{r_B}{6r_g}\right)^{-1/2}=\left(\frac{c_{s,\infty}^2+\Delta v^2}{6c^2}\right)^{1/2}, 
\end{align}
where $r_B\equiv G M_{\rm BH}/(c_{s,\infty}^2+\Delta v^2)$ is the Bondi radius, $r_g\equiv G M_{\rm BH}/c^2$ is the gravitational radius, and $\kappa_{\rm cho}$ is the suppression factor obtained in \cite{Cho+2024,Cho+2025}.

AGN feedback is treated as an isotropic wind that predominantly carries kinetic energy using the particle spawning method. The mass and energy fluxes of the outflow are given by
\begin{align} \dot{M}_{\rm wind}=f_{\rm mass, wind} \dot{M}_{\rm Bondi}.\notag\\ 
\dot{E}_{\rm wind} =\eta \dot{M}_{\rm BH} c^2= \frac{1}{2} \dot{M}_{\rm wind} V_{\rm wind} ^2, \end{align}
where $f_{\rm mass, wind}=1-\kappa_{\rm cho}$ for runs with feedback and $\eta$ is the efficiency of kinetic feedback. The corresponding wind velocity is given by:
\begin{align}\label{eq:vel}
V_{\rm wind} &= \left(\frac{2\eta f_{\rm mass, acc}}{f_{\rm mass, wind}}\right)^{1/2} c= \left(\frac{2\eta \kappa_{\rm cho}}{1-\kappa_{\rm cho}}\right)^{1/2} c.
\end{align}

Details of the numerical implementation are given in \aref{a:agn}. Limitations of the model, such as neglected geometric effects, are discussed further in \citet{Su+2025Bridging} and improvements are deferred to future work.

We perform an on-the-fly friend-of-friends search over all particles and seed a fixed-mass black hole once the stellar mass reaches a specified threshold. We consider three seeding scenarios: (1) \textit{Fiducial seeding} -- when the stellar mass reaches $3\times10^{10}M_\odot/h$, we seed a $3.5\times10^{7}M_\odot/h$ black hole; (2) \textit{Light seeding} -- at the same threshold, we seed a $3.5\times10^{6}M_\odot/h$ black hole; and (3) \textit{Early seeding} -- when the stellar mass reaches $3\times10^{9}M_\odot/h$, we seed a $3.5\times10^{7}M_\odot/h$ black hole. The first seeding events occur at $z=4.5$, 4.5, and 7.2, respectively. Once seeded, the black hole is gradually migrated toward the potential minimum via an additional artificial acceleration toward the most-bound collisionless particle in its interaction kernel each time-step \citep{Wellons2023}.

For each seeding model, we run a subset of five simulations: (i) Run A -- no or minimal accretion and no feedback; (ii) Run B -- black hole accretion suppressed relative to the Bondi rate ($f_{\rm mass, acc}=\kappa_{\rm Cho}$) but no AGN feedback; and (iii) Runs C1, C2, and C3 -- suppressed Bondi accretion with three AGN feedback efficiencies, $\eta = 0.02, 0.3, 1$. The first two values span the low to high efficiencies predicted by multi-zone GRMHD simulations \citep{Cho+2025} for $a_*=0$ and $a_*=0.9$. The third, $\eta=1$, corresponds to $a_*=0.9$ and steady prograde rotation \citep[e.g.,][]{Cho-Narayan2025}. Run parameters are summarized in \tref{tab:ic}.

\begin{table*}
\begin{center}
 \caption{Parameter Choices and Initial Conditions in our Simulations}
 \label{tab:ic}
 
\resizebox{18.cm}{!}
 {
  \movetableright=-1.3in
\begin{tabular}{ll|cc|cc|ccccc}

\hline
&&\multicolumn{2}{c|}{Numerical details}&\multicolumn{2}{c|}{Seeding parameter} & \multicolumn{5}{c}{Feedback parameter} \\
\hline
\multicolumn{2}{c|}{Model} & $z_{\rm final}$ & $m_{\rm wind}$  & $M_{\rm star}$ & $M_{\rm BH}$  &  $T_{\rm wind}$&  $B_{\rm wind}$  & $f_{\rm mass, acc}$ & $f_{\rm mass, wind}$ & $\eta_{\rm Kin}$ \\
&& & (${\rm M_\odot/h}$)& (${\rm M_\odot/h}$)& (${\rm M_\odot/h}$) & (K) & (G) &  &    \\
\hline
\multicolumn{2}{l|}{\underline{\textcolor{blue}{Fiducial Seeding}}}&&&&&\\
A/B&Fiducial--$\eta=0.$        &1& - & $3\times10^{10}$ & $3.5\times10^7$& -  & - & $\kappa_{\rm Cho}$ & $1-\kappa_{\rm Cho}$ &0 \\
C1&Fiducial--$\eta=0.02$        &1& $5\times10^4$ & $3\times10^{10}$  & $3.5\times10^7$&$10^7$  &$10^{-4}$& $\kappa_{\rm Cho}$ & $1-\kappa_{\rm Cho}$ &0.02\\
C2&Fiducial--$\eta=0.3$        &1& $5\times10^4$ & $3\times10^{10}$  & $3.5\times10^7$&$10^7$  &$10^{-4}$& $\kappa_{\rm Cho}$ & $1-\kappa_{\rm Cho}$ &0.3\\
C3&Fiducial--$\eta=1$        &1& $5\times10^4$ & $3\times10^{10}$  & $3.5\times10^7$&$10^7$  &$10^{-4}$& $\kappa_{\rm Cho}$ & $1-\kappa_{\rm Cho}$ &1\\
\hline
\multicolumn{2}{l|}{\underline{\textcolor{blue}{Light Seeding}}}&&&&&\\
B&Light--$\eta=0.$        &1& - & $3\times10^{10}$ & $3.5\times10^6$& -  & - & $\kappa_{\rm Cho}$ & $1-\kappa_{\rm Cho}$ &0 \\
C1&Light--$\eta=0.02$        &1& $5\times10^4$ & $3\times10^{10}$  & $3.5\times10^6$&$10^7$  &$10^{-4}$& $\kappa_{\rm Cho}$ & $1-\kappa_{\rm Cho}$ &0.02\\
C3&Light--$\eta=1$        &1& $5\times10^4$ & $3\times10^{10}$  & $3.5\times10^6$&$10^7$  &$10^{-4}$& $\kappa_{\rm Cho}$ & $1-\kappa_{\rm Cho}$ &1\\
\hline
\multicolumn{2}{l|}{\underline{\textcolor{blue}{Early Seeding}}}&&&&&\\
A&Early--NoAcc/NoFB        &1& - & $3\times10^{9}$ & $3.5\times10^7$& -  & - &0 & 0 &0 \\
B&Early--$\eta=0.$        &1& - & $3\times10^{9}$ & $3.5\times10^7$& -  & - & $\kappa_{\rm Cho}$ & $1-\kappa_{\rm Cho}$ &0 \\
C1&Early--$\eta=0.3$        &1& $5\times10^4$ & $3\times10^{9}$  & $3.5\times10^7$&$10^7$  &$10^{-4}$& $\kappa_{\rm Cho}$ & $1-\kappa_{\rm Cho}$ &0.3\\
C3&Early--$\eta=1$        &1& $5\times10^4$ & $3\times10^{9}$  & $3.5\times10^7$&$10^7$  &$10^{-4}$& $\kappa_{\rm Cho}$ & $1-\kappa_{\rm Cho}$ &1\\
\hline 
\hline
\end{tabular}
}
\end{center}
\begin{flushleft}

List of the runs: 
(1)Model name. 
(2) $z_{\rm final}$: Final redshift for the run. 
(3) $m_{\rm wind}$: Mass resolution of the spawned AGN wind particles. 
(4) $M_{\rm star}$: Stellar mass to initiate seeding. 
(5) $M_{\rm BH}$: Seed black hole mass.
(6,7) $T_{\rm wind}$ and $B_{\rm wind}$. 
(8,9) $f_{\rm mass, acc}$ $f_{\rm mass, wind}$: The accretion rate and the wind mass flux with respect to Bondi. 
(10) $\eta_{\rm Kin}$: The  kinetic AGN feedback efficiencies. Note that $\eta_{\rm Kin} = 0.02$ corresponds to a spin of $a_* = 0$, while $\eta_{\rm Kin} = 0.3$ corresponds to $a_* = 0.9$.
\end{flushleft}
\end{table*}

\section{Results}
\label{S:results}

\subsection{Properties of the Stellar and Black Hole Masses}\label{sec:everything}
\begin{figure*}
    \centering
    \includegraphics[width=\linewidth]{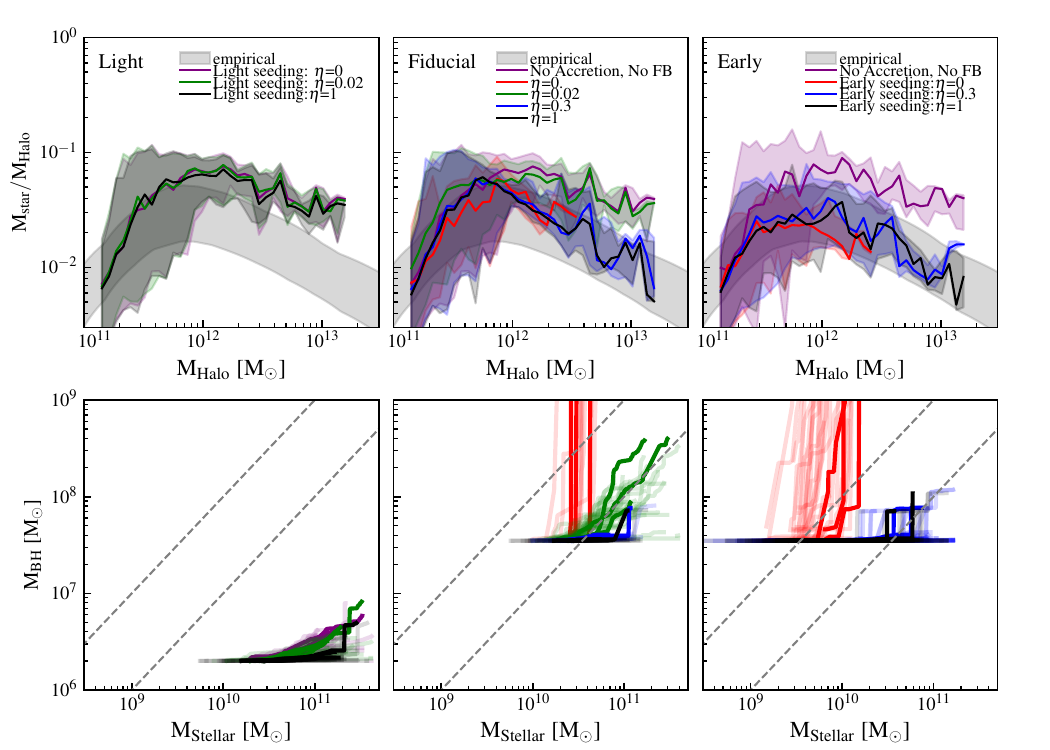}
    \caption{Stellar mass--halo mass relation (top) and black hole mass--stellar mass relation (bottom) for the light-seeding (left), fiducial-seeding (center), and early-seeding (right) models. The shaded region, which represents the empirical relation at $z = 0$, should be treated as a guide for the eye, as we overplot results from all redshifts.
For seed black holes with $M_\mathrm{BH}<10^7\,M_\odot$, the black hole does not grow, and the galaxy is not quenched regardless of feedback efficiencies, resulting in excess stellar mass above $M_{\rm Halo}\gtrsim 10^{12}M_\odot$. In the fiducial seeding case, the black hole overgrows without feedback, grows moderately when the efficiency matches $a_*=0$ result, and fails to accrete when the efficiency corresponds to maximum spin ($\eta \gtrsim 0.3$); star formation is also quenched in this high-efficiency case. With early massive seeding, black holes still do not grow through accretion for $\eta \gtrsim 0.3$, but growth through mergers can reproduce the observed local black hole--stellar mass relation. Galaxy quenching is also more efficient, yielding better agreement with the local stellar mass--halo mass relation. }
    \label{fig:statistical}
\end{figure*}

\fref{fig:statistical} presents the statistical properties of the black hole population, the host galaxy stellar masses, and corresponding halo masses in the simulations, shown in both the stellar mass--halo mass relation and the black hole mass--stellar mass relation. For simplicity, we plot the properties of all subhalos where the Lagrangian region is contaminated by $<10\%$ low-resolution elements, 
evaluated across all times. Different columns correspond to different seeding models.

For black hole seeds with $M_\mathrm{BH}<10^7\,M_\odot$ (light seeding), both accretion and the resulting feedback are weak. As a result, the black holes fail to grow and the star formation in galaxies also remain unquenched. This negligible growth persists even in the absence of feedback ($\eta=0$). Consequently, the stellar mass produced above $M_{\rm Halo}\gtrsim 10^{12}M_\odot$ consistently exceeds empirical values, due to the weak AGN feedback.  

In the fiducial seeding case, both accretion and feedback are stronger, and the feedback efficiency critically influences black hole growth. Without feedback, the black hole overgrows, while with efficiencies matching the $a_*=0$ result ($\eta= 0.02$) it grows only moderately. For efficiencies corresponding to the high spinning case studied ($\eta \geq 0.3$), the black hole fails to grow through accretion. Star formation is quenched only in this high-efficiency regime.  

We therefore encounter a intriguing dilemma: {\it the black hole either accretes without quenching star formation, as in the case of lower feedback efficiency for the case of a non-spinning ($a_*=0$) black hole, or it quenches star formation but fails to grow through accretion, as in the case of higher feedback efficiency for rapidly spinning black holes ($a_*>0.9$).}

With early massive seeding, a comparable mass black hole  begins accreting and providing feedback earlier at higher redshift. As before, black holes are seen to overgrow compared to the local relation in the absence of feedback, but fail to grow through accretion for $\eta \geq0.3$. The difference is that in the early seeding case, although it still does not accrete substantially, black hole growth through mergers may possibly reproduce the observed local black hole--stellar mass relation (see Section~\ref{sec:dilemma}).
The earlier presence of massive black holes also makes quenching more efficient, yielding better agreement with the local stellar--halo mass relation.

\begin{figure*}
    \centering
    \includegraphics[width=\linewidth]{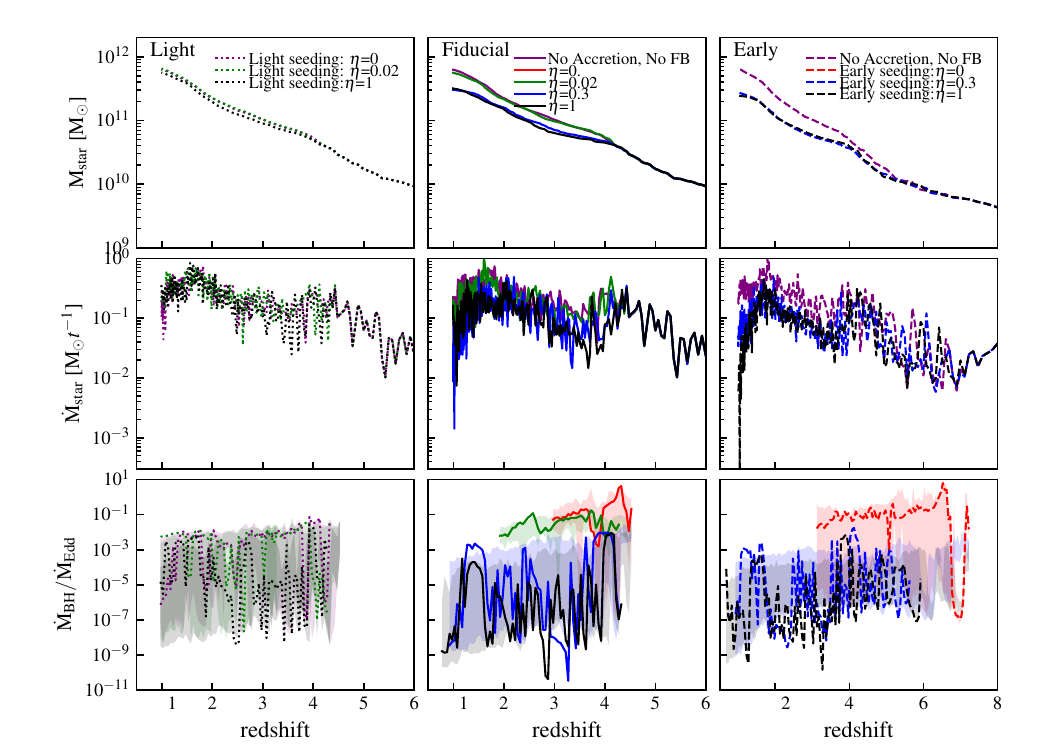}
    \caption{The stellar mass (top), star formation rate (middle), and black hole accretion rate (bottom) of the most massive progenitor in the light-seeding (left), fiducial-seeding (center), and early-seeding (right) models.
The shaded region shows the 16--84 percentile range of the accretion rate across all black holes in the run at each time. Black holes with $M_\mathrm{BH}\lesssim10^7\,M_\odot$ do not impact the star formation rate regardless of feedback efficiency. For $M_\mathrm{BH}\gtrsim3\times10^7\,M_\odot$, feedback efficiencies $\eta \gtrsim 0.3$ ($a_*\gtrsim0.9$) lead to efficient quenching of star formation, while efficiencies $\eta \lesssim 0.02$ remain ineffective. A feedback efficiency $\gtrsim0.3$ always suppresses black hole accretion, regardless of black hole mass.}
    \label{fig:massive}
\end{figure*}

In addition to the statistical properties, a similar conclusion can be drawn from \fref{fig:massive}, which shows the stellar mass, star formation rate, and black hole accretion rate of the most massive progenitor as a function of time. In addition to the accretion rate of the specific black hole, the shaded region indicates the 16--84 th percentile distribution of the BH population.  

In order to suppress star formation in the most massive progenitor, the black hole must exceed $3\times10^7\,M_\odot/h$ and have a feedback efficiency $\geq0.3$ (corresponding to a spin $a_*\gtrsim0.9$). Otherwise, the feedback remains too weak to be impactful.  

Similarly, suppressing black hole growth requires a feedback efficiency of $\eta\geq0.3$ (or $a_*\gtrsim0.9$) to produce a visible effect.

\subsection{Suppression of accretion}

\begin{figure*}
    \centering
    \includegraphics[width=\linewidth]{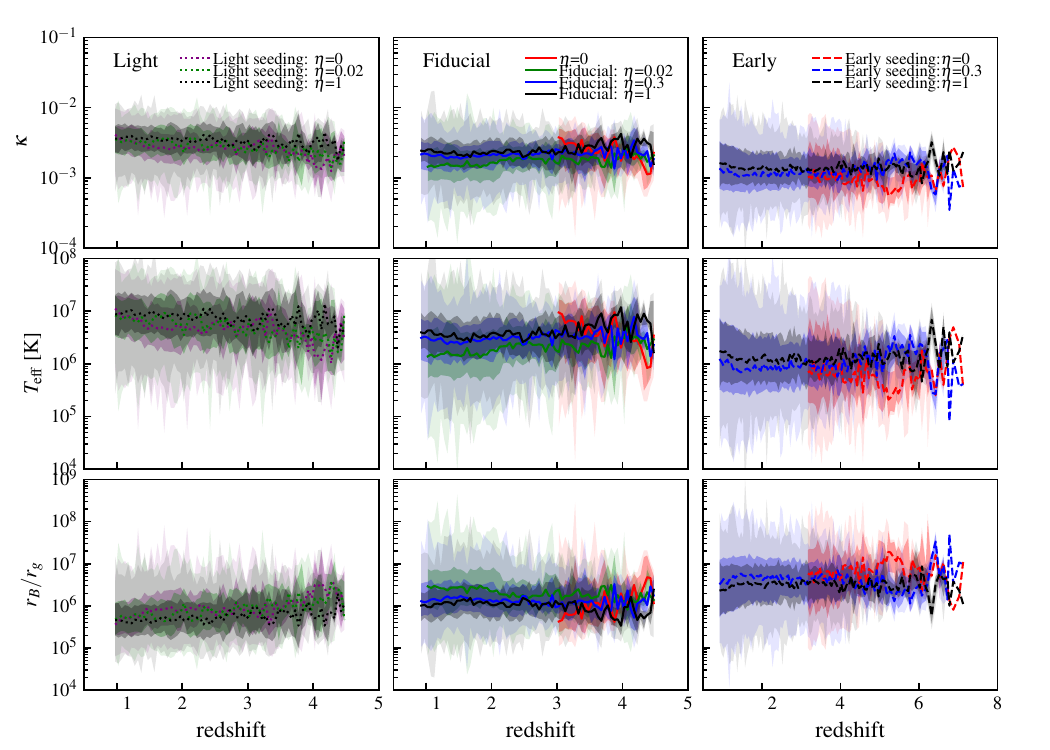}
    \caption{The evolution of accretion suppression (top, $\kappa \equiv \dot{M}_{\rm BH}/\dot{M}_{\rm Bondi}$), the corresponding gas temperature (middle), and the Bondi radius (bottom) for the light-seeding (left), fiducial-seeding (center), and early-seeding (right) models.
The shaded regions show the full distribution and the 16--84 percentile range. In all cases, the suppression of black hole accretion relative to the Bondi rate falls within $\sim10^{-3}$--$10^{-2}$. Higher feedback efficiency produces hotter gas around the black hole and thus larger $\kappa$. More massive black holes, resulting from earlier or more massive seeding, lead to broader distributions of $\kappa$, temperature, and $r_B/r_g$, and systematically lower average $\kappa$ and temperatures.  }
    \label{fig:kappa}
\end{figure*}

The dilemma manifests itself as the system falling into one of two regimes: the black hole either accretes without quenching or quenches but does not grow. This outcome arises from the strong suppression of the accretion rate relative to the Bondi rate.
\fref{fig:kappa} shows the time evolution of $\kappa$, together with the corresponding effective temperature around the black hole and the implied Bondi radius relative to the gravitational radius (\Eqref{eq:kaappa_cho}). The shaded regions denote the 16--84 th percentile range as well as the full distribution at each time.  Note that the effective temperature is defined as the average $T_{\rm eff} = m(c_s^2 + \Delta v^2)/\gamma k$ within the black hole kernel, where $m$ is the mean particle mass.

Regardless of the seeding model, higher feedback efficiency consistently produces hotter gas near the black hole and a larger separation between $r_B$ and $r_g$, which corresponds to a larger $\kappa$. In all cases, $\kappa$ lies within $10^{-3}$--$10^{-2}$, with a median value of $1$--$3\times10^{-3}$. This represents a dramatic suppression and is the primary reason why it is difficult for black holes to grow while simultaneously quenching star formation.  

We also find that more massive black holes, arising from either earlier or heavier seeding, produce broader distributions of $\kappa$, temperature, and $r_B/r_g$. This occurs because a more massive black hole can drive stronger episodes of feedback, temporarily heating the surrounding gas, while also encountering denser gas due to the deeper potential well. The net effect is that, despite the broader distribution, the median $\kappa$ for more massive black holes -- whether due to earlier or more massive seeding -- is lower, reflecting a lower ambient temperature.

\subsection{Gas morphology}
\begin{figure*}
    \centering
    \includegraphics[width=\linewidth]{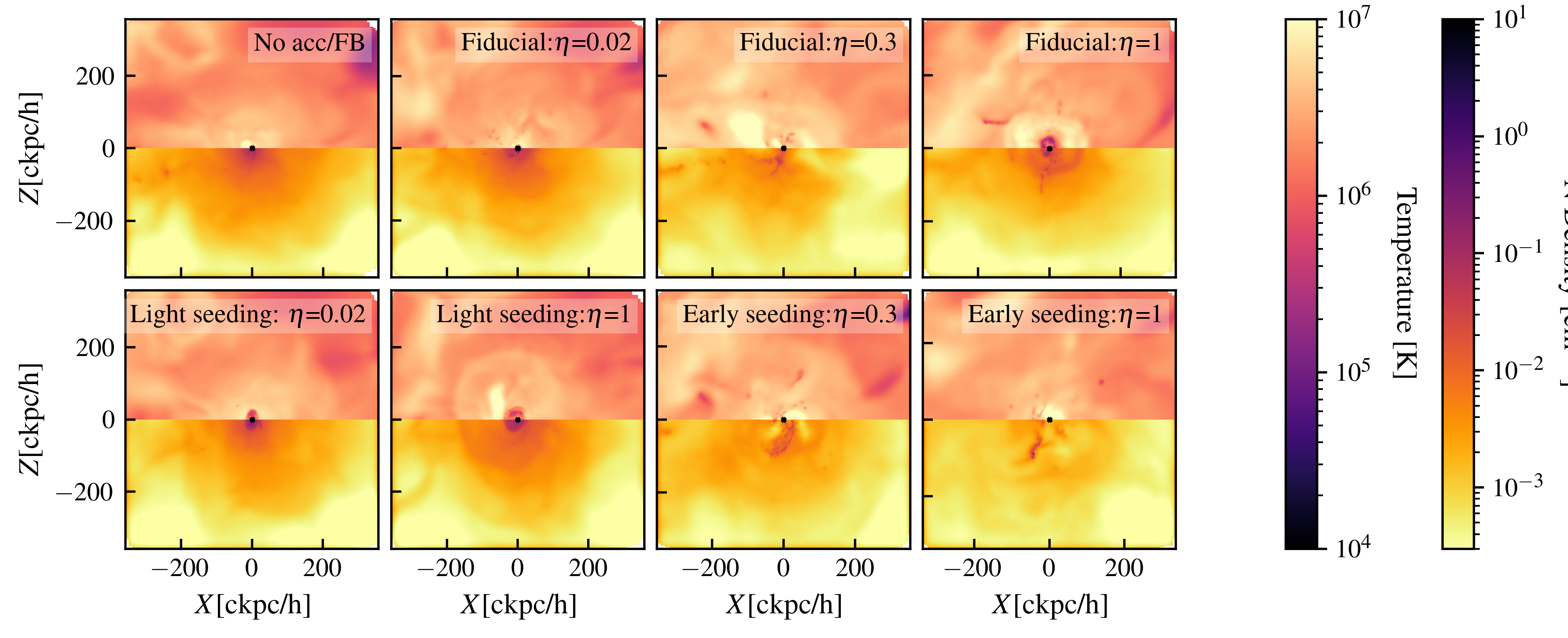}
    \caption{The gas temperature (upper half) and density (lower half) morphology at $z\sim1.2$ for seven runs with AGN feedback and one without.
For $\eta = 0.02$, AGN feedback produces minimal changes to the gas morphology, while efficiencies $\eta \gtrsim 0.3$ generate visible outflows extending to large radii. 
}
\label{fig:morphology}
\end{figure*}

\fref{fig:morphology} shows the gas temperature and density morphology at $z\sim1.2$ for the seven runs with non-zero feedback efficiency, along with one run without feedback for reference. Because the timing of individual AGN feedback episodes varies from run to run, we selected snapshots that best illustrate the differences. Runs with higher feedback efficiency ($\eta\geq0.3$) inject feedback with a higher specific energy, producing stronger episodes that propagate to larger radii. This is evident in both the higher temperatures and the reduced densities that extend to large scales.

\section{Discussion and Conclusions}
\label{S:conclusions}
\subsection{Paths Out of the Dilemma}\label{sec:dilemma}
In \sref{S:results}, we showed that, if multi-zone GRMHD--informed magnetically suppressed black hole accretion and AGN feedback are assumed to be the only modes of growth and feedback, the system faces the following dilemma: the black hole either accretes without quenching star formation, as in the case of lower feedback efficiency or a non-spinning ($a_*=0$) black hole, or it quenches but fails to grow through accretion, as in the case of higher feedback efficiency or rapidly spinning black holes ($a_*\gtrsim0.9$). Moreover, the black hole does not grow through this mode of accretion before reaching $\sim10^7\,M_\odot/h$, even in the absence of feedback. We also noted that this dilemma originates from the highly suppressed accretion rate. Yet, in the universe, we clearly observe black holes more massive than $10^8\,M_\odot$ as well as quenched galaxies. 
To reconcile this dilemma with the observations, there are several possible ways forward.

\begin{itemize}
\item{\bf A separate cold or super-Eddington accretion mode:\,\,\,}
The multi-zone GRMHD simulations are designed to mimic an SMBH in a hot, sub-Eddington environment, where accretion is strongly suppressed by magnetic fields. However, the mode of black hole growth informed by these simulations may not be the only relevant scenario -- particularly at earlier stages, before the black hole has grown into an SMBH and before the host galaxy has evolved into a massive cluster elliptical.  A separate treatment of cold gas accretion, such as via gravitational torque models \cite{Hopkins2010,AngleAlcazar+2017,AngleAlcazar+2017_2,Bryne2023} or chaotic cold accretion models \citep{Gaspari+2013}, may be required to explain black hole growth and feedback in the supermassive regime, and appears essential for describing black hole growth prior to reaching $\sim10^7\,M_\odot/h$.  In such models, a larger fraction of the available gas can be funneled onto the black hole (see also how gravitational instabilities contribute to BH mass assembly in Dark Sage, \citealt{Porras+2025_Dark_Sage}). This pathway is also consistent with observations of cold gas around black holes \citep[e.g.,][]{Gorski+2025}, as well as with previous studies concluding that black holes do not gain significant mass during the ADAF phase, and that most of their growth must have occurred during an earlier, radiatively efficient, super-Eddington stage \citep{Hopkins_Narayan2006, Hopkins+2006_merger_driven_acc}.
Part of this issue is explored in our separate work using semi-analytic models (SAM) in Porras et al. 2025c (submitted).
\item{\bf Spin dependence of feedback:\,\,\,}
We demonstrate that for a feedback efficiency of $\eta\sim0.02$, mimicking zero black hole spin, the black hole grows but does not quench, while for $\eta\geq0.3$, corresponding to the rapidly spinning case of the multi-zone GRMHD simulations, the opposite occurs. A possible way forward is to track black hole spin by following the angular momentum of the accreted gas, and including spin-down due to feedback. For example, incoherent accretion could keep spins low during periods of efficient accretion, while BH-BH mergers could provide larger spins at later times \citep{Ricarte+2025}. This might possibly allow a balance between the two extremes: during low-spin episodes the black hole can grow, while during high-spin episodes it can produce strong feedback and quench star formation. 
\item{\bf Collimated feedback:\,\,\,} Although we model our feedback as isotropic regardless of black hole spin, jet feedback from a rapidly spinning black hole remains collimated at the Bondi radius \citep{Cho+2025}. When the outflow geometry is properly accounted for, such collimated feedback may be less effective at suppressing accretion near the black hole than the isotropic case with identical total feedback efficiency. As a result, it can sustain a more stable accretion flow while depositing energy at larger radii, balancing black hole growth and galaxy quenching. This will be left for future study and is further discussed in \aref{a:geometry}.

\item{\bf Abundant early massive black holes boost merger-driven growth:\,\,\,}
We partially demonstrate this with runs adopting an earlier, more massive seeding criterion: when a halo’s stellar mass first reaches $3\times10^9\,M_\odot/h$, we initiate a black hole with $M_{\rm BH}/M_\star\sim1/100$. In this case, although the black holes still do not accrete for high feedback efficiencies ($\eta\gtrsim0.3$), the looser seeding threshold substantially increases the number of black holes and hence the merger rate. This enhanced merger activity is sufficient to evolve the system from $M_{\rm BH}/M_\star\sim1/100$ to the local relation of $\sim1/1000$. These runs agree best with the local black hole--stellar mass relation as well as with empirical stellar mass--halo mass relations.
\end{itemize}

Either one or a combination of the four paths outlined above should be sufficient to reconcile the discrepancy between the dilemma and the observational evidence. We emphasize, however, that the latter three paths can only remedy the situation once a black hole has already reached a few times $10^7\,M_\odot$, while a separate treatment of cold or super-Eddington accretion is essential for growing the black hole to that mass scale in the first place.

\subsection{Summary}

\begin{itemize}

\item \textbf{Bridging scales:}  
We present a unified framework that connects the physics of black hole accretion and feedback across the full hierarchy of scales--from the event horizon to the galactic environment--by embedding a GRMHD-informed, multi-zone accretion and feedback model \citep{Cho+2023,Cho+2024,Cho+2025} within cosmological magnetohydrodynamic zoom-in simulations.

\item \textbf{First-principles modeling:}  
This approach links relativistic inflows and outflows to galaxy-scale dynamics, enabling a self-consistent exploration of how supermassive black holes grow and regulate their hosts in the suppressed Bondi regime.

\item \textbf{Key finding -- suppressed growth:}  
Multi-zone GRMHD-informed feedback yields accretion rates far lower than predicted by traditional Bondi models, effectively stalling SMBH growth until black holes exceed masses of $\sim10^{7}\,M_\odot$ even without any feedback.

\item \textbf{Bifurcation in evolutionary outcomes:}  For black holes above $\sim10^{7}\,M_\odot$, the system falls into the following two cases under the multi-zone GRMHD-informed feedback and accretion model.
\begin{itemize}
    \item Low-spin systems ($a_*\sim0$ or $\eta \sim 0.02$): Continue to accrete at low rates, without quenching star formation.
    \item High-spin systems ($a_*=0.9$ or $\eta \gtrsim 0.3$): Quench star formation efficiently but rapidly deplete their gas supply, limiting further black hole growth.
\end{itemize}

\item \textbf{Physical implications:}  
This spin-dependent bifurcation explains the observed coexistence of massive quenched galaxies with dormant black holes and star-forming systems with modestly growing nuclei.

\item \textbf{Need for additional channels:}  
GRMHD-regulated accretion alone cannot account for the full observed SMBH mass spectrum. Additional growth mechanisms--major mergers, cold inflows, or brief super-Eddington episodes--are required, particularly at high redshift.

Our results highlight the need for an evolving, multi-regime model of SMBH growth that captures the interplay between spin, feedback, and environment, offering a physically grounded framework for co-evolution across cosmic time.

\end{itemize}

\section*{Acknowledgements}

 KS, PN, HC, RN, and BP were partially supported by the Black Hole Initiative at Harvard University, which is funded by the Gordon and Betty Moore Foundation grant 8273, and the John Templeton Foundation grant 61497. The opinions expressed in this publication are those of the authors and do not necessarily reflect the views of the Moore or Templeton Foundation. CAFG was supported by NSF through grants AST-2108230 and AST-2307327; by NASA through grants 21-ATP21-0036 and 23-ATP23-0008; and by STScI through grant JWST-AR-03252.001-A. Support for PFH was provided by NSF Research Grants 1911233, 20009234, 2108318, NSF CAREER grant 1455342, NASA grants 80NSSC18K0562, HST-AR-15800.   
This work was performed in part at Aspen Center for Physics, which is supported by National Science Foundation grant PHY-2210452. The simulations are done on Frontera with allocation AST22010, and Bridges-2 with Access allocations TG-PHY220027 \& TGPHY220047. Support for PFH was provided by NSF Research Grants 1911233, 20009234, 2108318, NSF CA-REER grant 1455342, NASA grants 80NSSC18K0562, HST-AR-15800.   This research also used resources provided to BP by the Los Alamos National Laboratory Institutional Computing Program, which is supported by the U.S. Department of Energy National Nuclear Security Administration under Contract No. 89233218CNA000001.  We thank FIRE and LtU collaboration for useful discussions.

\appendix
\section{Numerical details}
\subsection{Stellar Physics}\label{a:sfb}
Our simulations utilize the FIRE-2 (Feedback In Realistic Environments) framework for modeling the interstellar medium (ISM), star formation, and stellar feedback, as described in detail in \citet{Hopkins+FIRE2018}, which also presents extensive numerical validations. Radiative cooling is followed from $10$ to $10^{10}\,\mathrm{K}$, accounting for a wide range of processes including photoelectric and photoionization heating, Compton and collisional cooling, fine-structure emission, recombination, and both atomic and molecular cooling channels. The simulations implement a time-dependent but uniform cosmic UV/X-ray background following \cite{Faucher-Giguere2009}.

Star formation occurs in gas that is dense ($n > 1000\,\mathrm{cm^{-3}}$), molecular, self-shielding, and locally self-gravitating, implemented through a sink-particle scheme \citep{Hopkins+SFC2013}. Once formed, star particles represent entire stellar populations with metallicities inherited from their progenitor gas particles. Stellar feedback rates--including supernovae, stellar winds, and radiation--are computed using IMF-averaged outputs from {\small STARBURST99} \citep{Leitherer+Starburst99_1999}, assuming a \citet{Kroupa+IMF2002} initial mass function.

The feedback model includes several key processes:  
(1) radiative feedback from photoionization, photoelectric heating, and radiation pressure across five wavelength bands (ionizing, FUV, NUV, optical--NIR, and IR);  
(2) stellar winds from OB and AGB stars, which contribute continuous injection of mass, metals, momentum, and thermal energy; and  
(3) both Type Ia and Type II supernovae, including prompt and delayed events, with ejecta and energetics prescribed from tabulated stellar evolution models. 
Magnetohydrodynamics (MHD) is included in all runs.

\subsection{ BH Accretion and the Density and temperature estimate within}\label{a:acc}
To evaluate the gas density used in the Bondi accretion rate and the suppression factor $\kappa_{\rm Cho}$, we compute kernel-weighted averages of the density and temperature within a radius $r_d$. This radius is defined as enclosing 96 kernel-weighted neighbors, with a minimum enforced value of $r_d > 50$ pc/h. In the light seeding runs, we instead adopt a directly averaged density over the same radius; however, we find this has a negligible impact on the resulting black hole accretion, in part because these black holes do not accrete significantly under the assumed seeding conditions.

Spawned gas particles are explicitly excluded from the density and temperature calculations to prevent noise in the estimated local quantities, and they are also not allowed to be accreted by the black hole.

As shown in \citet{Cho+2024} (see their Fig. 5), the gas density profile is relatively flat beyond the Bondi radius over the range $10^5$--$10^8\,r_g$ (corresponding to $\sim$30 pc--10 kpc). For simplicity, we therefore adopt the density and temperature values measured directly in the simulation without applying any rescaling. A more detailed analysis of the implications of this choice is provided in \citet{Su+2025Bridging}.

We also note that the Bondi accretion formula used in \Eqref{eq:bondi} for suppressed Bondi accretion is the canonical Bondi-Hoyle-Lyttleton expression, which accounts for the relative velocity between the gas and the black hole. However, the corresponding velocity dependence is not verified by multi-zone GRMHD simulations. Velocity dispersion is also not accounted for in the expression. We apply the same formula to all gas phases, including low-temperature gas, assuming the functional dependence is universally applicable, although multi-zone GRMHD simulations have only tested it in the regime of $T_{\infty} \gtrsim 10^6$ K. The broader applicability will be left for future multi-zone GRMHD work. The expression also asymptotes to 4 times the adiabatic Bondi accretion rate assuming $\gamma = 5/3$, and is numerically closer to the case assuming $\gamma = 1$, given that galaxy simulations are non-adiabatic and such an expression behaves better in the limit where the sound speed is low. We emphasize that uncertainties in the overall normalization--by a factor of a few--arising from either density estimation or the choice of Bondi expression, and which apply simultaneously to both the black hole accretion rate and wind mass flux, should not change our modeling or qualitative conclusions. Following \Eqref{eq:vel}, the jet velocity, which determines the specific energy of the jet, depends only on $\eta$ and $\kappa_{\rm cho}$. Moreover, from \cite{Su+2025Bridging}, we also conclude that under similar settings, feedback models self-regulate to a similar average AGN feedback energy $\langle\dot{E}_{\rm wind}\rangle$ \citep[see also][]{Su+2023,Su+2025+box}, and the implied black hole accretion rate $\langle\dot{m}_{\rm BH}\rangle \sim \langle\dot{E}_{\rm wind}\rangle/\eta c^2$ depends only on the feedback efficiency $\eta$. If anything, if the overall Bondi accretion rate is further lowered by a factor of a few following adiabatic expression, the black hole accretion rate may be reduced even more, making the dilemma in \sref{sec:dilemma} slightly more severe.

\subsection{AGN Wind Implementation}\label{a:agn}

We implement AGN wind launching using a particle spawning method \citep{Torrey+2020,Su+2021,Weinberger+2023,Su+JetHaloMass2024}, which injects new gas resolution elements directly from the central black hole (BH). This method offers improved control over jet properties, as it reduces sensitivity to neighbor-finding routines and allows us to enforce higher resolution for the launched material. The feedback setup used here is similar to that in \citet{Torrey+2020}, which explored the impact of broad absorption line (BAL) winds on disk galaxies.

The mass resolution of spawned gas particles is listed in \tref{tab:ic}, and they are prevented from de-refining (i.e., merging into standard gas elements) until their velocity drops below 10\% of the launch speed. To conserve linear momentum exactly, we inject two particles simultaneously in opposite directions whenever the accumulated jet mass flux equals twice the target mass per particle.

Each wind particle is initialized at a random point on a sphere of radius $r_0$, defined as either 50 comoving pc/h or half the BH--nearest-gas-particle separation, whichever is smaller. The initial velocity vector is radial. The magnetic field of the launched particle is initialized to $<10^{-4}\,\mathrm{G}$ in the $z$-direction, and the initial gas temperature is set to $10^7\,\mathrm{K}$. Both magnetic and thermal energy fluxes are subdominant compared to the kinetic energy flux of the wind.

\subsection{Initial conditions}\label{a:ic}
The initial conditions consist of a $100$ comoving Mpc/$h$ box with a zoom-in region centered on a halo that reaches $\sim10^{14}\,M_\odot$ at $z=0$. Within the zoom-in region, the dark-matter particle mass is $8.4\times10^6\,M_\odot/h$, and the initial gas mass resolution is $1.6\times10^6\,M_\odot/h$. The initial seeded magnetic field is set to $10^{-10}$ G.
The run was first evolved without black holes to $z = 7.2$ for the early seed case, or to $z = 4.5$ for all other runs listed in \tref{tab:ic}, at which point the simulations were restarted with black holes and AGN feedback, corresponding to the earliest seeding time implied by each black hole seeding model.

Once a black hole is seeded, we enable on-the-fly super-Lagrangian hyperrefinement of gas particles, achieving a maximum resolution of $4\times10^5\,M_\odot/h$ within $15$ comoving kpc/$h$ of the BH. All runs are evolved to $z=1$ (and not $z=0$, because of computational cost); by this epoch the principal differences among models are already apparent. A full evolution to $z=0$ is deferred to future work.

\section{Possible future improvements and current limitations}\label{a:numerical}
\subsection{{   Geometrical effects and efficiency of the AGN feedback}}\label{a:geometry}

Although \citet{Cho+2025} obtain an AGN jet for the non-zero black hole spin case, we launch an isotropic wind in this study for simplicity; this is a key limitation of the present study. At our resolution, geometric effects of AGN feedback may be subdominant. Moreover, in the higher specific-energy cases ($\eta \gtrsim 0.3$)--which correspond to higher black hole spins--the initially elongated jet cocoon tends to isotropize and evolve into a quasi-spherical bubble on small scales, even when it is adequately resolved \citep{Su+2021,Su+JetHaloMass2024}.

However, an initially more collimated jet may still be less efficient at regulating the cooling flow and quenching star formation and black hole growth \citep{Su+2021,Su+JetHaloMass2024}. Meanwhile, a more collimated feedback may sustain steadier black hole accretion while transporting energy to larger radii.  The impact of such geometric effects and a detailed modeling of the angular dependence of wind mass, kinetic, and thermal energy fluxes are left to future exploration.

\subsection{Resolution effect}\label{a:geometry}
The gas mass resolution in this study is $\sim 4\times 10^5\,M_\odot$, which is sufficient to ensure convergence for SNe feedback\citep{Hopkins+2018SNe} and a good behair for FIRE stellar feedback models \cite[e.g.,][]{2017MNRAS.470.1050F}. However, it is lower than that in our previous isolated galaxy simulations \citep{Su+2025Bridging}, due to the higher computational cost of cosmological zoom-in runs.

The implications for black hole accretion and AGN feedback remain broadly consistent with the discussion in \citet{Su+2025Bridging}. The dominant uncertainties still lie in the estimation of the accretion rate, rather than in the feedback injection process. The resolution of the spawned wind particles differs by only a factor of $\sim 3.5$ compared to our previous isolated runs. For the feedback injection itself, resolution effects are expected to become more significant in future work if we explicitly model the angular distribution of jet mass and energy fluxes.

Regarding black hole accretion, the Bondi radius is often unresolved in the vicinity of the black hole, particularly when the black hole mass is still relatively low. Nevertheless, we estimate the Bondi accretion rate and the suppression prefactor ($\kappa_{\rm Cho}$) using kernel-weighted averages of gas density and temperature within a smoothing radius, without applying any rescaling despite the radius of the kernel. This choice is motivated by the multi-zone GRMHD simulations of \citet{Cho+2024}, which include realistic galactic potentials and gas profiles, and demonstrate that the density and temperature remain relatively flat over the range $10^5$--$10^8\,r_g$ (corresponding to $\sim$30\,pc--3\,kpc). 
We therefore assume that the values measured in our simulations are representative of those at the Bondi radius, had it been resolved.

Given this, we do not expect our resolution to qualitatively bias the overall black hole accretion rate. However, we acknowledge that limited resolution may introduce scatter in the accretion rate over time. A more detailed examination of these effects is deferred to future work.

\subsection{Black hole positioning and mergers}\label{a:geometry}
At the current resolution, it is not possible to accurately capture the dynamical friction required to sink the black hole to the potential minimum. We therefore adopt the approach of \citet{Wellons2023}, in which the black hole identifies the most deeply bound particle and is repositioned continuously with a velocity-damping term. This combination yields the most stable centering behavior in practice. However, this subgrid repositioning scheme may affect the black hole merger rate, and its impact will be explored in future studies with higher resolution.

\bibliography{bridging_scales}{}

\begin{thebibliography}{}
\expandafter\ifx\csname natexlab\endcsname\relax\def\natexlab#1{#1}\fi
\providecommand{\url}[1]{\href{#1}{#1}}
\providecommand{\dodoi}[1]{doi:~\href{http://doi.org/#1}{\nolinkurl{#1}}}
\providecommand{\doeprint}[1]{\href{http://ascl.net/#1}{\nolinkurl{http://ascl.net/#1}}}
\providecommand{\doarXiv}[1]{\href{https://arxiv.org/abs/#1}{\nolinkurl{https://arxiv.org/abs/#1}}}

\bibitem[{{Angl{\'e}s-Alc{\'a}zar}
  {et~al.}(2017{\natexlab{a}}){Angl{\'e}s-Alc{\'a}zar}, {Dav{\'e}},
  {Faucher-Gigu{\`e}re}, {{\"O}zel}, \& {Hopkins}}]{AngleAlcazar+2017}
{Angl{\'e}s-Alc{\'a}zar}, D., {Dav{\'e}}, R., {Faucher-Gigu{\`e}re}, C.-A.,
  {{\"O}zel}, F., \& {Hopkins}, P.~F. 2017{\natexlab{a}}, \mnras, 464, 2840,
  \dodoi{10.1093/mnras/stw2565}

\bibitem[{{Angl{\'e}s-Alc{\'a}zar}
  {et~al.}(2017{\natexlab{b}}){Angl{\'e}s-Alc{\'a}zar}, {Faucher-Gigu{\`e}re},
  {Quataert}, {Hopkins}, {Feldmann}, {Torrey}, {Wetzel}, \&
  {Kere{\v{s}}}}]{AngleAlcazar+2017_2}
{Angl{\'e}s-Alc{\'a}zar}, D., {Faucher-Gigu{\`e}re}, C.-A., {Quataert}, E.,
  {et~al.} 2017{\natexlab{b}}, \mnras, 472, L109, \dodoi{10.1093/mnrasl/slx161}

\bibitem[{{Angl{\'e}s-Alc{\'a}zar} {et~al.}(2021){Angl{\'e}s-Alc{\'a}zar},
  {Quataert}, {Hopkins}, {Somerville}, {Hayward}, {Faucher-Gigu{\`e}re},
  {Bryan}, {Kere{\v{s}}}, {Hernquist}, \& {Stone}}]{AnglesAlcazar+Zoom_2021}
{Angl{\'e}s-Alc{\'a}zar}, D., {Quataert}, E., {Hopkins}, P.~F., {et~al.} 2021,
  \apj, 917, 53, \dodoi{10.3847/1538-4357/ac09e8}

\bibitem[{{Baldry} {et~al.}(2004){Baldry}, {Glazebrook}, {Brinkmann},
  {Ivezi{\'c}}, {Lupton}, {Nichol}, \& {Szalay}}]{Baldry2004}
{Baldry}, I.~K., {Glazebrook}, K., {Brinkmann}, J., {et~al.} 2004, \apj, 600,
  681, \dodoi{10.1086/380092}

\bibitem[{{Bell} {et~al.}(2003){Bell}, {McIntosh}, {Katz}, \&
  {Weinberg}}]{Bell+2003}
{Bell}, E.~F., {McIntosh}, D.~H., {Katz}, N., \& {Weinberg}, M.~D. 2003, \apjs,
  149, 289, \dodoi{10.1086/378847}

\bibitem[{{Blanton} {et~al.}(2005){Blanton}, {Eisenstein}, {Hogg}, {Schlegel},
  \& {Brinkmann}}]{Blanton+2005}
{Blanton}, M.~R., {Eisenstein}, D., {Hogg}, D.~W., {Schlegel}, D.~J., \&
  {Brinkmann}, J. 2005, \apj, 629, 143, \dodoi{10.1086/422897}

\bibitem[{{Byrne} {et~al.}(2023){Byrne}, {Faucher-Gigu{\`e}re}, {Stern},
  {Angl{\'e}s-Alc{\'a}zar}, {Wellons}, {Gurvich}, \& {Hopkins}}]{Bryne2023}
{Byrne}, L., {Faucher-Gigu{\`e}re}, C.-A., {Stern}, J., {et~al.} 2023, \mnras,
  520, 722, \dodoi{10.1093/mnras/stad171}

\bibitem[{{Byrne} {et~al.}(2024){Byrne}, {Faucher-Gigu{\`e}re}, {Wellons},
  {Hopkins}, {Angl{\'e}s-Alc{\'a}zar}, {Sultan}, {Wijers}, {Moreno}, \&
  {Ponnada}}]{2024ApJ...973..149B}
{Byrne}, L., {Faucher-Gigu{\`e}re}, C.-A., {Wellons}, S., {et~al.} 2024, \apj,
  973, 149, \dodoi{10.3847/1538-4357/ad67ca}

\bibitem[{{Chatterjee} {et~al.}(2023){Chatterjee}, {Chael}, {Tiede}, {Mizuno},
  {Emami}, {Fromm}, {Ricarte}, {Blackburn}, {Roelofs}, {Johnson}, {Doeleman},
  {Arras}, {Fuentes}, {Knollm{\"u}ller}, {Kosogorov}, {Lindahl}, {M{\"u}ller},
  {Patel}, {Raymond}, {Traianou}, \& {Vega}}]{Chatterjee:2023}
{Chatterjee}, K., {Chael}, A., {Tiede}, P., {et~al.} 2023, Galaxies, 11, 38,
  \dodoi{10.3390/galaxies11020038}

\bibitem[{{Cho} \& {Narayan}(2025)}]{Cho-Narayan2025}
{Cho}, H., \& {Narayan}, R. 2025, \apj, 991, 89,
  \dodoi{10.3847/1538-4357/adf8d3}

\bibitem[{{Cho} {et~al.}(2023){Cho}, {Prather}, {Narayan}, {Natarajan}, {Su},
  {Ricarte}, \& {Chatterjee}}]{Cho+2023}
{Cho}, H., {Prather}, B.~S., {Narayan}, R., {et~al.} 2023, \apjl, 959, L22,
  \dodoi{10.3847/2041-8213/ad1048}

\bibitem[{{Cho} {et~al.}(2025){Cho}, {Prather}, {Narayan}, {Su}, \&
  {Natarajan}}]{Cho+2025}
{Cho}, H., {Prather}, B.~S., {Narayan}, R., {Su}, K.-Y., \& {Natarajan}, P.
  2025, arXiv e-prints, arXiv:2507.17818, \dodoi{10.48550/arXiv.2507.17818}

\bibitem[{{Cho} {et~al.}(2024){Cho}, {Prather}, {Su}, {Narayan}, \&
  {Natarajan}}]{Cho+2024}
{Cho}, H., {Prather}, B.~S., {Su}, K.-Y., {Narayan}, R., \& {Natarajan}, P.
  2024, arXiv e-prints, arXiv:2405.13887, \dodoi{10.48550/arXiv.2405.13887}

\bibitem[{{Dekel} \& {Birnboim}(2006)}]{Dekel+2006}
{Dekel}, A., \& {Birnboim}, Y. 2006, \mnras, 368, 2,
  \dodoi{10.1111/j.1365-2966.2006.10145.x}

\bibitem[{{Event Horizon Telescope Collaboration} {et~al.}(2021){Event Horizon
  Telescope Collaboration}, {Akiyama}, {Algaba}, {Alberdi}, {Alef}, {Anantua},
  {Asada}, {Azulay}, {Baczko}, {Ball}, {Balokovi{\'c}}, {Barrett}, {Benson},
  {Bintley}, {Blackburn}, {Blundell}, {Boland}, {Bouman}, {Bower}, {Boyce},
  {Bremer}, {Brinkerink}, {Brissenden}, {Britzen}, {Broderick}, {Broguiere},
  {Bronzwaer}, {Byun}, {Carlstrom}, {Chael}, {Chan}, {Chatterjee},
  {Chatterjee}, {Chen}, {Chen}, {Chesler}, {Cho}, {Christian}, {Conway},
  {Cordes}, {Crawford}, {Crew}, {Cruz-Osorio}, {Cui}, {Davelaar}, {De
  Laurentis}, {Deane}, {Dempsey}, {Desvignes}, {Dexter}, {Doeleman}, {Eatough},
  {Falcke}, {Farah}, {Fish}, {Fomalont}, {Ford}, {Fraga-Encinas}, {Friberg},
  {Fromm}, {Fuentes}, {Galison}, {Gammie}, {Garc{\'\i}a}, {Gelles}, {Gentaz},
  {Georgiev}, {Goddi}, {Gold}, {G{\'o}mez}, {G{\'o}mez-Ruiz}, {Gu}, {Gurwell},
  {Hada}, {Haggard}, {Hecht}, {Hesper}, {Himwich}, {Ho}, {Ho}, {Honma},
  {Huang}, {Huang}, {Hughes}, {Ikeda}, {Inoue}, {Issaoun}, {James}, {Jannuzi},
  {Janssen}, {Jeter}, {Jiang}, {Jimenez-Rosales}, {Johnson}, {Jorstad}, {Jung},
  {Karami}, {Karuppusamy}, {Kawashima}, {Keating}, {Kettenis}, {Kim}, {Kim},
  {Kim}, {Kim}, {Kino}, {Koay}, {Kofuji}, {Koch}, {Koyama}, {Kramer}, {Kramer},
  {Krichbaum}, {Kuo}, {Lauer}, {Lee}, {Levis}, {Li}, {Li}, {Lindqvist}, {Lico},
  {Lindahl}, {Liu}, {Liu}, {Liuzzo}, {Lo}, {Lobanov}, {Loinard}, {Lonsdale},
  {Lu}, {MacDonald}, {Mao}, {Marchili}, {Markoff}, {Marrone}, {Marscher},
  {Mart{\'\i}-Vidal}, {Matsushita}, {Matthews}, {Medeiros}, {Menten}, {Mizuno},
  {Mizuno}, {Moran}, {Moriyama}, {Moscibrodzka}, {M{\"u}ller}, {Musoke}, {Mus
  Mej{\'\i}as}, {Michalik}, {Nadolski}, {Nagai}, {Nagar}, {Nakamura},
  {Narayan}, {Narayanan}, {Natarajan}, {Nathanail}, {Neilsen}, {Neri}, {Ni},
  {Noutsos}, {Nowak}, {Okino}, {Olivares}, {Ortiz-Le{\'o}n}, {Oyama},
  {{\"O}zel}, {Palumbo}, {Park}, {Patel}, {Pen}, {Pesce}, {Pi{\'e}tu},
  {Plambeck}, {PopStefanija}, {Porth}, {P{\"o}tzl}, {Prather},
  {Preciado-L{\'o}pez}, {Psaltis}, {Pu}, {Ramakrishnan}, {Rao}, {Rawlings},
  {Raymond}, {Rezzolla}, {Ricarte}, {Ripperda}, {Roelofs}, {Rogers}, {Ros},
  {Rose}, {Roshanineshat}, {Rottmann}, {Roy}, {Ruszczyk}, {Rygl},
  {S{\'a}nchez}, {S{\'a}nchez-Arguelles}, {Sasada}, {Savolainen}, {Schloerb},
  {Schuster}, {Shao}, {Shen}, {Small}, {Sohn}, {SooHoo}, {Sun}, {Tazaki},
  {Tetarenko}, {Tiede}, {Tilanus}, {Titus}, {Toma}, {Torne}, {Trent},
  {Traianou}, {Trippe}, {van Bemmel}, {van Langevelde}, {van Rossum}, {Wagner},
  {Ward-Thompson}, {Wardle}, {Weintroub}, {Wex}, {Wharton}, {Wielgus}, {Wong},
  {Wu}, {Yoon}, {Young}, {Young}, {Younsi}, {Yuan}, {Yuan}, {Zensus}, {Zhao},
  \& {Zhao}}]{EHTm87_8_2021}
{Event Horizon Telescope Collaboration}, {Akiyama}, K., {Algaba}, J.~C.,
  {et~al.} 2021, \apjl, 910, L13, \dodoi{10.3847/2041-8213/abe4de}

\bibitem[{{Event Horizon Telescope Collaboration} {et~al.}(2022){Event Horizon
  Telescope Collaboration}, {Akiyama}, {Alberdi}, {Alef}, {Algaba}, {Anantua},
  {Asada}, {Azulay}, {Bach}, {Baczko}, {Ball}, {Balokovi{\'c}}, {Barrett},
  {Baub{\"o}ck}, {Benson}, {Bintley}, {Blackburn}, {Blundell}, {Bouman},
  {Bower}, {Boyce}, {Bremer}, {Brinkerink}, {Brissenden}, {Britzen},
  {Broderick}, {Broguiere}, {Bronzwaer}, {Bustamante}, {Byun}, {Carlstrom},
  {Ceccobello}, {Chael}, {Chan}, {Chatterjee}, {Chatterjee}, {Chen}, {Chen},
  {Cheng}, {Cho}, {Christian}, {Conroy}, {Conway}, {Cordes}, {Crawford},
  {Crew}, {Cruz-Osorio}, {Cui}, {Davelaar}, {De Laurentis}, {Deane}, {Dempsey},
  {Desvignes}, {Dexter}, {Dhruv}, {Doeleman}, {Dougal}, {Dzib}, {Eatough},
  {Emami}, {Falcke}, {Farah}, {Fish}, {Fomalont}, {Ford}, {Fraga-Encinas},
  {Freeman}, {Friberg}, {Fromm}, {Fuentes}, {Galison}, {Gammie}, {Garc{\'\i}a},
  {Gentaz}, {Georgiev}, {Goddi}, {Gold}, {G{\'o}mez-Ruiz}, {G{\'o}mez}, {Gu},
  {Gurwell}, {Hada}, {Haggard}, {Haworth}, {Hecht}, {Hesper}, {Heumann}, {Ho},
  {Ho}, {Honma}, {Huang}, {Huang}, {Hughes}, {Ikeda}, {Violette Impellizzeri},
  {Inoue}, {Issaoun}, {James}, {Jannuzi}, {Janssen}, {Jeter}, {Jiang},
  {Jim{\'e}nez-Rosales}, {Johnson}, {Jorstad}, {Joshi}, {Jung}, {Karami},
  {Karuppusamy}, {Kawashima}, {Keating}, {Kettenis}, {Kim}, {Kim}, {Kim},
  {Kim}, {Kino}, {Koay}, {Kocherlakota}, {Kofuji}, {Koch}, {Koyama}, {Kramer},
  {Kramer}, {Krichbaum}, {Kuo}, {La Bella}, {Lauer}, {Lee}, {Lee}, {Leung},
  {Levis}, {Li}, {Lico}, {Lindahl}, {Lindqvist}, {Lisakov}, {Liu}, {Liu},
  {Liuzzo}, {Lo}, {Lobanov}, {Loinard}, {Lonsdale}, {Lu}, {Mao}, {Marchili},
  {Markoff}, {Marrone}, {Marscher}, {Mart{\'\i}-Vidal}, {Matsushita},
  {Matthews}, {Medeiros}, {Menten}, {Michalik}, {Mizuno}, {Mizuno}, {Moran},
  {Moriyama}, {Moscibrodzka}, {M{\"u}ller}, {Mus}, {Musoke}, {Myserlis},
  {Nadolski}, {Nagai}, {Nagar}, {Nakamura}, {Narayan}, {Narayanan},
  {Natarajan}, {Nathanail}, {Navarro Fuentes}, {Neilsen}, {Neri}, {Ni},
  {Noutsos}, {Nowak}, {Oh}, {Okino}, {Olivares}, {Ortiz-Le{\'o}n}, {Oyama},
  {{\"O}zel}, {Palumbo}, {Filippos Paraschos}, {Park}, {Parsons}, {Patel},
  {Pen}, {Pesce}, {Pi{\'e}tu}, {Plambeck}, {PopStefanija}, {Porth},
  {P{\"o}tzl}, {Prather}, {Preciado-L{\'o}pez}, {Psaltis}, {Pu},
  {Ramakrishnan}, {Rao}, {Rawlings}, {Raymond}, {Rezzolla}, {Ricarte},
  {Ripperda}, {Roelofs}, {Rogers}, {Ros}, {Romero-Ca{\~n}izales},
  {Roshanineshat}, {Rottmann}, {Roy}, {Ruiz}, {Ruszczyk}, {Rygl},
  {S{\'a}nchez}, {S{\'a}nchez-Arg{\"u}elles}, {S{\'a}nchez-Portal}, {Sasada},
  {Satapathy}, {Savolainen}, {Schloerb}, {Schonfeld}, {Schuster}, {Shao},
  {Shen}, {Small}, {Sohn}, {SooHoo}, {Souccar}, {Sun}, {Tazaki}, {Tetarenko},
  {Tiede}, {Tilanus}, {Titus}, {Torne}, {Traianou}, {Trent}, {Trippe}, {Turk},
  {van Bemmel}, {van Langevelde}, {van Rossum}, {Vos}, {Wagner},
  {Ward-Thompson}, {Wardle}, {Weintroub}, {Wex}, {Wharton}, {Wielgus}, {Wiik},
  {Witzel}, {Wondrak}, {Wong}, {Wu}, {Yamaguchi}, {Yoon}, {Young}, {Young},
  {Younsi}, {Yuan}, {Yuan}, {Zensus}, {Zhang}, {Zhao}, {Zhao}, {Chan}, {Qiu},
  {Ressler}, \& {White}}]{EHTMW_5_2022}
{Event Horizon Telescope Collaboration}, {Akiyama}, K., {Alberdi}, A., {et~al.}
  2022, \apjl, 930, L16, \dodoi{10.3847/2041-8213/ac6672}

\bibitem[{{Fabian} {et~al.}(1994){Fabian}, {Arnaud}, {Bautz}, \&
  {Tawara}}]{Fabian+1994}
{Fabian}, A.~C., {Arnaud}, K.~A., {Bautz}, M.~W., \& {Tawara}, Y. 1994, \apjl,
  436, L63, \dodoi{10.1086/187633}

\bibitem[{{Faucher-Gigu{\`e}re} {et~al.}(2009){Faucher-Gigu{\`e}re}, {Lidz},
  {Zaldarriaga}, \& {Hernquist}}]{Faucher-Giguere2009}
{Faucher-Gigu{\`e}re}, C.-A., {Lidz}, A., {Zaldarriaga}, M., \& {Hernquist}, L.
  2009, \apj, 703, 1416, \dodoi{10.1088/0004-637X/703/2/1416}

\bibitem[{{Feldmann} \& {Mayer}(2015)}]{Feldmann+2015}
{Feldmann}, R., \& {Mayer}, L. 2015, \mnras, 446, 1939,
  \dodoi{10.1093/mnras/stu2207}

\bibitem[{{Feldmann} {et~al.}(2017){Feldmann}, {Quataert}, {Hopkins},
  {Faucher-Gigu{\`e}re}, \& {Kere{\v{s}}}}]{2017MNRAS.470.1050F}
{Feldmann}, R., {Quataert}, E., {Hopkins}, P.~F., {Faucher-Gigu{\`e}re}, C.-A.,
  \& {Kere{\v{s}}}, D. 2017, \mnras, 470, 1050, \dodoi{10.1093/mnras/stx1120}

\bibitem[{{Fiacconi} {et~al.}(2018){Fiacconi}, {Sijacki}, \&
  {Pringle}}]{Fiacconi2018}
{Fiacconi}, D., {Sijacki}, D., \& {Pringle}, J.~E. 2018, \mnras, 477, 3807,
  \dodoi{10.1093/mnras/sty893}

\bibitem[{{Gammie} {et~al.}(2003){Gammie}, {McKinney}, \&
  {T{\'o}th}}]{Gammie2003}
{Gammie}, C.~F., {McKinney}, J.~C., \& {T{\'o}th}, G. 2003, \apj, 589, 444,
  \dodoi{10.1086/374594}

\bibitem[{{Gaspari} {et~al.}(2013){Gaspari}, {Ruszkowski}, \&
  {Oh}}]{Gaspari+2013}
{Gaspari}, M., {Ruszkowski}, M., \& {Oh}, S.~P. 2013, \mnras, 432, 3401,
  \dodoi{10.1093/mnras/stt692}

\bibitem[{{Gorski} \& {Murchikova}(2025)}]{Gorski+2025}
{Gorski}, M.~D., \& {Murchikova}, E. 2025, arXiv e-prints, arXiv:2509.10615,
  \dodoi{10.48550/arXiv.2509.10615}

\bibitem[{{Guo} {et~al.}(2023){Guo}, {Stone}, {Kim}, \& {Quataert}}]{Guo2023}
{Guo}, M., {Stone}, J.~M., {Kim}, C.-G., \& {Quataert}, E. 2023, \apj, 946, 26,
  \dodoi{10.3847/1538-4357/acb81e}

\bibitem[{{Guo} {et~al.}(2024){Guo}, {Stone}, {Quataert}, \& {Kim}}]{Guo+2024}
{Guo}, M., {Stone}, J.~M., {Quataert}, E., \& {Kim}, C.-G. 2024, arXiv
  e-prints, arXiv:2405.11711, \dodoi{10.48550/arXiv.2405.11711}

\bibitem[{{Guo} {et~al.}(2025){Guo}, {Stone}, {Quataert}, \&
  {Springel}}]{Guo+2025}
{Guo}, M., {Stone}, J.~M., {Quataert}, E., \& {Springel}, V. 2025, \apj, 987,
  202, \dodoi{10.3847/1538-4357/add1da}

\bibitem[{{Hopkins}(2015)}]{Hopkins2015mfm}
{Hopkins}, P.~F. 2015, \mnras, 450, 53, \dodoi{10.1093/mnras/stv195}

\bibitem[{{Hopkins}(2016)}]{Hopkins+2015mhd_divergence}
---. 2016, \mnras, 462, 576, \dodoi{10.1093/mnras/stw1578}

\bibitem[{{Hopkins} {et~al.}(2006{\natexlab{a}}){Hopkins}, {Hernquist}, {Cox},
  {Di Matteo}, {Robertson}, \& {Springel}}]{Hopkins+2006_merger_driven_acc}
{Hopkins}, P.~F., {Hernquist}, L., {Cox}, T.~J., {et~al.} 2006{\natexlab{a}},
  \apjs, 163, 1, \dodoi{10.1086/499298}

\bibitem[{{Hopkins} \& {Most}(2025)}]{Hopkins+2025timedialation}
{Hopkins}, P.~F., \& {Most}, E.~R. 2025, arXiv e-prints, arXiv:2510.09756,
  \dodoi{10.48550/arXiv.2510.09756}

\bibitem[{{Hopkins} {et~al.}(2006{\natexlab{b}}){Hopkins}, {Narayan}, \&
  {Hernquist}}]{Hopkins_Narayan2006}
{Hopkins}, P.~F., {Narayan}, R., \& {Hernquist}, L. 2006{\natexlab{b}}, \apj,
  643, 641, \dodoi{10.1086/503154}

\bibitem[{{Hopkins} {et~al.}(2013){Hopkins}, {Narayanan}, \&
  {Murray}}]{Hopkins+SFC2013}
{Hopkins}, P.~F., {Narayanan}, D., \& {Murray}, N. 2013, \mnras, 432, 2647,
  \dodoi{10.1093/mnras/stt723}

\bibitem[{{Hopkins} \& {Quataert}(2010)}]{Hopkins2010}
{Hopkins}, P.~F., \& {Quataert}, E. 2010, \mnras, 407, 1529,
  \dodoi{10.1111/j.1365-2966.2010.17064.x}

\bibitem[{{Hopkins} \& {Raives}(2016)}]{Hopkins+2016mhd}
{Hopkins}, P.~F., \& {Raives}, M.~J. 2016, \mnras, 455, 51,
  \dodoi{10.1093/mnras/stv2180}

\bibitem[{{Hopkins} {et~al.}(2018{\natexlab{a}}){Hopkins}, {Wetzel},
  {Kere{\v{s}}}, {Faucher-Gigu{\`e}re}, {Quataert}, {Boylan-Kolchin}, {Murray},
  {Hayward}, {Garrison-Kimmel}, {Hummels}, {Feldmann}, {Torrey}, {Ma},
  {Angl{\'e}s-Alc{\'a}zar}, {Su}, {Orr}, {Schmitz}, {Escala}, {Sanderson},
  {Grudi{\'c}}, {Hafen}, {Kim}, {Fitts}, {Bullock}, {Wheeler}, {Chan},
  {Elbert}, \& {Narayanan}}]{Hopkins+FIRE2018}
{Hopkins}, P.~F., {Wetzel}, A., {Kere{\v{s}}}, D., {et~al.} 2018{\natexlab{a}},
  \mnras, 480, 800, \dodoi{10.1093/mnras/sty1690}

\bibitem[{{Hopkins} {et~al.}(2018{\natexlab{b}}){Hopkins}, {Wetzel},
  {Kere{\v{s}}}, {Faucher-Gigu{\`e}re}, {Quataert}, {Boylan-Kolchin}, {Murray},
  {Hayward}, \& {El-Badry}}]{Hopkins+2018SNe}
---. 2018{\natexlab{b}}, \mnras, 477, 1578, \dodoi{10.1093/mnras/sty674}

\bibitem[{{Hopkins} {et~al.}(2023){Hopkins}, {Grudic}, {Su}, {Wellons},
  {Angles-Alcazar}, {Steinwandel}, {Guszejnov}, {Murray}, {Faucher-Giguere},
  {Quataert}, \& {Keres}}]{Hopkinsforgedinfire2023}
{Hopkins}, P.~F., {Grudic}, M.~Y., {Su}, K.-Y., {et~al.} 2023, arXiv e-prints,
  arXiv:2309.13115.
\newblock \doarXiv{2309.13115}

\bibitem[{{Kaaz} {et~al.}(2024){Kaaz}, {Liska}, {Tchekhovskoy}, {Hopkins}, \&
  {Jacquemin-Ide}}]{Kaaz2024}
{Kaaz}, N., {Liska}, M., {Tchekhovskoy}, A., {Hopkins}, P.~F., \&
  {Jacquemin-Ide}, J. 2024, arXiv e-prints, arXiv:2410.01877,
  \dodoi{10.48550/arXiv.2410.01877}

\bibitem[{{Kauffmann} {et~al.}(2003){Kauffmann}, {Heckman}, {White}, {Charlot},
  {Tremonti}, {Peng}, {Seibert}, {Brinkmann}, {Nichol}, {SubbaRao}, \&
  {York}}]{Kauffmann+2003}
{Kauffmann}, G., {Heckman}, T.~M., {White}, S.~D.~M., {et~al.} 2003, \mnras,
  341, 54, \dodoi{10.1046/j.1365-8711.2003.06292.x}

\bibitem[{{Kere{\v s}} {et~al.}(2009){Kere{\v s}}, {Katz}, {Dav{\'e}},
  {Fardal}, \& {Weinberg}}]{keres+2009}
{Kere{\v s}}, D., {Katz}, N., {Dav{\'e}}, R., {Fardal}, M., \& {Weinberg},
  D.~H. 2009, \mnras, 396, 2332, \dodoi{10.1111/j.1365-2966.2009.14924.x}

\bibitem[{{Kere{\v s}} {et~al.}(2005){Kere{\v s}}, {Katz}, {Weinberg}, \&
  {Dav{\'e}}}]{keres+2005}
{Kere{\v s}}, D., {Katz}, N., {Weinberg}, D.~H., \& {Dav{\'e}}, R. 2005,
  \mnras, 363, 2, \dodoi{10.1111/j.1365-2966.2005.09451.x}

\bibitem[{{Kroupa}(2002)}]{Kroupa+IMF2002}
{Kroupa}, P. 2002, Science, 295, 82, \dodoi{10.1126/science.1067524}

\bibitem[{{Lalakos} {et~al.}(2022){Lalakos}, {Gottlieb}, {Kaaz}, {Chatterjee},
  {Liska}, {Christie}, {Tchekhovskoy}, {Zhuravleva}, \&
  {Nokhrina}}]{Lalakos2022}
{Lalakos}, A., {Gottlieb}, O., {Kaaz}, N., {et~al.} 2022, \apjl, 936, L5,
  \dodoi{10.3847/2041-8213/ac7bed}

\bibitem[{{Leitherer} {et~al.}(1999){Leitherer}, {Schaerer}, {Goldader},
  {Delgado}, {Robert}, {Kune}, {de Mello}, {Devost}, \&
  {Heckman}}]{Leitherer+Starburst99_1999}
{Leitherer}, C., {Schaerer}, D., {Goldader}, J.~D., {et~al.} 1999, \apjs, 123,
  3, \dodoi{10.1086/313233}

\bibitem[{{Li} \& {Bryan}(2014)}]{Li2014}
{Li}, Y., \& {Bryan}, G.~L. 2014, \apj, 789, 153,
  \dodoi{10.1088/0004-637X/789/2/153}

\bibitem[{{Madgwick} {et~al.}(2003){Madgwick}, {Somerville}, {Lahav}, \&
  {Ellis}}]{Madgwick+2003}
{Madgwick}, D.~S., {Somerville}, R., {Lahav}, O., \& {Ellis}, R. 2003, \mnras,
  343, 871, \dodoi{10.1046/j.1365-8711.2003.06729.x}

\bibitem[{{McDonald} {et~al.}(2011){McDonald}, {Veilleux}, \&
  {Mushotzky}}]{McDonald+2011}
{McDonald}, M., {Veilleux}, S., \& {Mushotzky}, R. 2011, \apj, 731, 33,
  \dodoi{10.1088/0004-637X/731/1/33}

\bibitem[{{McDonald} {et~al.}(2012){McDonald}, {Bayliss}, {Benson}, {Foley},
  {Ruel}, {Sullivan}, {Veilleux}, {Aird}, {Ashby}, {Bautz}, {Bazin}, {Bleem},
  {Brodwin}, {Carlstrom}, {Chang}, {Cho}, {Clocchiatti}, {Crawford}, {Crites},
  {de Haan}, {Desai}, {Dobbs}, {Dudley}, {Egami}, {Forman}, {Garmire},
  {George}, {Gladders}, {Gonzalez}, {Halverson}, {Harrington}, {High},
  {Holder}, {Holzapfel}, {Hoover}, {Hrubes}, {Jones}, {Joy}, {Keisler}, {Knox},
  {Lee}, {Leitch}, {Liu}, {Lueker}, {Luong-van}, {Mantz}, {Marrone}, {McMahon},
  {Mehl}, {Meyer}, {Miller}, {Mocanu}, {Mohr}, {Montroy}, {Murray}, {Natoli},
  {Padin}, {Plagge}, {Pryke}, {Rawle}, {Reichardt}, {Rest}, {Rex}, {Ruhl},
  {Saliwanchik}, {Saro}, {Sayre}, {Schaffer}, {Shaw}, {Shirokoff}, {Simcoe},
  {Song}, {Spieler}, {Stalder}, {Staniszewski}, {Stark}, {Story}, {Stubbs},
  {{\v{S}}uhada}, {van Engelen}, {Vanderlinde}, {Vieira}, {Vikhlinin},
  {Williamson}, {Zahn}, \& {Zenteno}}]{McDonald+2012}
{McDonald}, M., {Bayliss}, M., {Benson}, B.~A., {et~al.} 2012, \nat, 488, 349,
  \dodoi{10.1038/nature11379}

\bibitem[{{Mercedes-Feliz} {et~al.}(2023){Mercedes-Feliz},
  {Angl{\'e}s-Alc{\'a}zar}, {Hayward}, {Cochrane}, {Terrazas}, {Wellons},
  {Richings}, {Faucher-Gigu{\`e}re}, {Moreno}, {Su}, {Hopkins}, {Quataert}, \&
  {Kere{\v{s}}}}]{Mercedes+2023}
{Mercedes-Feliz}, J., {Angl{\'e}s-Alc{\'a}zar}, D., {Hayward}, C.~C., {et~al.}
  2023, \mnras, 524, 3446, \dodoi{10.1093/mnras/stad2079}

\bibitem[{{Ni} {et~al.}(2022){Ni}, {Di Matteo}, {Bird}, {Croft}, {Feng},
  {Chen}, {Tremmel}, {DeGraf}, \& {Li}}]{Ni2022}
{Ni}, Y., {Di Matteo}, T., {Bird}, S., {et~al.} 2022, \mnras, 513, 670,
  \dodoi{10.1093/mnras/stac351}

\bibitem[{{O'Dea} {et~al.}(2008){O'Dea}, {Baum}, {Privon}, {Noel-Storr},
  {Quillen}, {Zufelt}, {Park}, {Edge}, {Russell}, {Fabian}, {Donahue},
  {Sarazin}, {McNamara}, {Bregman}, \& {Egami}}]{ODea+2008}
{O'Dea}, C.~P., {Baum}, S.~A., {Privon}, G., {et~al.} 2008, \apj, 681, 1035,
  \dodoi{10.1086/588212}

\bibitem[{{Peterson} \& {Fabian}(2006)}]{Peterson+2006}
{Peterson}, J.~R., \& {Fabian}, A.~C. 2006, \physrep, 427, 1,
  \dodoi{10.1016/j.physrep.2005.12.007}

\bibitem[{{Porras-Valverde} {et~al.}(2025){Porras-Valverde}, {Ricarte},
  {Natarajan}, {Somerville}, {Gabrielpillai}, \&
  {Yung}}]{Porras+2025_Dark_Sage}
{Porras-Valverde}, A.~J., {Ricarte}, A., {Natarajan}, P., {et~al.} 2025, arXiv
  e-prints, arXiv:2504.11566, \dodoi{10.48550/arXiv.2504.11566}

\bibitem[{{Porth} {et~al.}(2019){Porth}, {Chatterjee}, {Narayan}, {Gammie},
  {Mizuno}, {Anninos}, {Baker}, {Bugli}, {Chan}, {Davelaar}, {Del Zanna},
  {Etienne}, {Fragile}, {Kelly}, {Liska}, {Markoff}, {McKinney}, {Mishra},
  {Noble}, {Olivares}, {Prather}, {Rezzolla}, {Ryan}, {Stone}, {Tomei},
  {White}, {Younsi}, {Akiyama}, {Alberdi}, {Alef}, {Asada}, {Azulay}, {Baczko},
  {Ball}, {Balokovi{\'c}}, {Barrett}, {Bintley}, {Blackburn}, {Boland},
  {Bouman}, {Bower}, {Bremer}, {Brinkerink}, {Brissenden}, {Britzen},
  {Broderick}, {Broguiere}, {Bronzwaer}, {Byun}, {Carlstrom}, {Chael},
  {Chatterjee}, {Chen}, {Chen}, {Cho}, {Christian}, {Conway}, {Cordes},
  {Geoffrey}, {Crew}, {Cui}, {De Laurentis}, {Deane}, {Dempsey}, {Desvignes},
  {Doeleman}, {Eatough}, {Falcke}, {Fish}, {Fomalont}, {Fraga-Encinas},
  {Freeman}, {Friberg}, {Fromm}, {G{\'o}mez}, {Galison}, {Garc{\'\i}a},
  {Gentaz}, {Georgiev}, {Goddi}, {Gold}, {Gu}, {Gurwell}, {Hada}, {Hecht},
  {Hesper}, {Ho}, {Ho}, {Honma}, {Huang}, {Huang}, {Hughes}, {Ikeda}, {Inoue},
  {Issaoun}, {James}, {Jannuzi}, {Janssen}, {Jeter}, {Jiang}, {Johnson},
  {Jorstad}, {Jung}, {Karami}, {Karuppusamy}, {Kawashima}, {Keating},
  {Kettenis}, {Kim}, {Kim}, {Kim}, {Kino}, {Koay}, {Patrick}, {Koch}, {Koyama},
  {Kramer}, {Kramer}, {Krichbaum}, {Kuo}, {Lauer}, {Lee}, {Li}, {Li},
  {Lindqvist}, {Liu}, {Liuzzo}, {Lo}, {Lobanov}, {Loinard}, {Lonsdale}, {Lu},
  {MacDonald}, {Mao}, {Marrone}, {Marscher}, {Mart{\'\i}-Vidal}, {Matsushita},
  {Matthews}, {Medeiros}, {Menten}, {Mizuno}, {Moran}, {Moriyama},
  {Moscibrodzka}, {M{\"u}ller}, {Nagai}, {Nagar}, {Nakamura}, {Narayanan},
  {Natarajan}, {Neri}, {Ni}, {Noutsos}, {Okino}, {Oyama}, {{\"O}zel},
  {Palumbo}, {Patel}, {Pen}, {Pesce}, {Pi{\'e}tu}, {Plambeck}, {PopStefanija},
  {Preciado-L{\'o}pez}, {Psaltis}, {Pu}, {Ramakrishnan}, {Rao}, {Rawlings},
  {Raymond}, {Ripperda}, {Roelofs}, {Rogers}, {Ros}, {Rose}, {Roshanineshat},
  {Rottmann}, {Roy}, {Ruszczyk}, {Rygl}, {S{\'a}nchez},
  {S{\'a}nchez-Arguelles}, {Sasada}, {Savolainen}, {Schloerb}, {Schuster},
  {Shao}, {Shen}, {Small}, {Sohn}, {SooHoo}, {Tazaki}, {Tiede}, {Tilanus},
  {Titus}, {Toma}, {Torne}, {Trent}, {Trippe}, {Tsuda}, {van Bemmel}, {van
  Langevelde}, {van Rossum}, {Wagner}, {Wardle}, {Weintroub}, {Wex}, {Wharton},
  {Wielgus}, {Wong}, {Wu}, {Young}, {Young}, {Yuan}, {Yuan}, {Zensus}, {Zhao},
  {Zhao}, {Zhu}, \& {Event Horizon Telescope Collaboration}}]{Porth:2019}
{Porth}, O., {Chatterjee}, K., {Narayan}, R., {et~al.} 2019, \apjs, 243, 26,
  \dodoi{10.3847/1538-4365/ab29fd}

\bibitem[{{Pozzetti} {et~al.}(2010){Pozzetti}, {Bolzonella}, {Zucca},
  {Zamorani}, {Lilly}, {Renzini}, {Moresco}, {Mignoli}, {Cassata}, {Tasca},
  {Lamareille}, {Maier}, {Meneux}, {Halliday}, {Oesch}, {Vergani}, {Caputi},
  {Kova{\v c}}, {Cimatti}, {Cucciati}, {Iovino}, {Peng}, {Carollo}, {Contini},
  {Kneib}, {Le F{\'e}vre}, {Mainieri}, {Scodeggio}, {Bardelli}, {Bongiorno},
  {Coppa}, {de la Torre}, {de Ravel}, {Franzetti}, {Garilli}, {Kampczyk},
  {Knobel}, {Le Borgne}, {Le Brun}, {Pell{\`o}}, {Perez Montero},
  {Ricciardelli}, {Silverman}, {Tanaka}, {Tresse}, {Abbas}, {Bottini}, {Cappi},
  {Guzzo}, {Koekemoer}, {Leauthaud}, {Maccagni}, {Marinoni}, {McCracken},
  {Memeo}, {Porciani}, {Scaramella}, {Scarlata}, \& {Scoville}}]{Pozzetti+2010}
{Pozzetti}, L., {Bolzonella}, M., {Zucca}, E., {et~al.} 2010, \aap, 523, A13,
  \dodoi{10.1051/0004-6361/200913020}

\bibitem[{{Rafferty} {et~al.}(2008){Rafferty}, {McNamara}, \&
  {Nulsen}}]{Rafferty+2008}
{Rafferty}, D.~A., {McNamara}, B.~R., \& {Nulsen}, P.~E.~J. 2008, \apj, 687,
  899, \dodoi{10.1086/591240}

\bibitem[{{Ressler} {et~al.}(2020){Ressler}, {White}, {Quataert}, \&
  {Stone}}]{Ressler2020}
{Ressler}, S.~M., {White}, C.~J., {Quataert}, E., \& {Stone}, J.~M. 2020,
  \apjl, 896, L6, \dodoi{10.3847/2041-8213/ab9532}

\bibitem[{{Ricarte} {et~al.}(2025){Ricarte}, {Natarajan}, {Narayan}, \&
  {Palumbo}}]{Ricarte+2025}
{Ricarte}, A., {Natarajan}, P., {Narayan}, R., \& {Palumbo}, D. C.~M. 2025,
  \apj, 980, 136, \dodoi{10.3847/1538-4357/ad9ea9}

\bibitem[{{Ricarte} {et~al.}(2019){Ricarte}, {Tremmel}, {Natarajan}, \&
  {Quinn}}]{Ricarte2019}
{Ricarte}, A., {Tremmel}, M., {Natarajan}, P., \& {Quinn}, T. 2019, \mnras,
  489, 802, \dodoi{10.1093/mnras/stz2161}

\bibitem[{{Rosas-Guevara} {et~al.}(2016){Rosas-Guevara}, {Bower}, {Schaye},
  {McAlpine}, {Dalla Vecchia}, {Frenk}, {Schaller}, \&
  {Theuns}}]{Rosas-Guevara2016}
{Rosas-Guevara}, Y., {Bower}, R.~G., {Schaye}, J., {et~al.} 2016, \mnras, 462,
  190, \dodoi{10.1093/mnras/stw1679}

\bibitem[{{Sijacki} {et~al.}(2015){Sijacki}, {Vogelsberger}, {Genel},
  {Springel}, {Torrey}, {Snyder}, {Nelson}, \& {Hernquist}}]{Sijacki2015}
{Sijacki}, D., {Vogelsberger}, M., {Genel}, S., {et~al.} 2015, \mnras, 452,
  575, \dodoi{10.1093/mnras/stv1340}

\bibitem[{{Stern} {et~al.}(2019){Stern}, {Fielding}, {Faucher-Gigu{\`e}re}, \&
  {Quataert}}]{Stern+2019}
{Stern}, J., {Fielding}, D., {Faucher-Gigu{\`e}re}, C.-A., \& {Quataert}, E.
  2019, \mnras, 488, 2549, \dodoi{10.1093/mnras/stz1859}

\bibitem[{{Su} {et~al.}(2025{\natexlab{a}}){Su}, {Bryan}, \&
  {Haiman}}]{Su+2025+box}
{Su}, K.-Y., {Bryan}, G.~L., \& {Haiman}, Z. 2025{\natexlab{a}}, \mnras, 538,
  11, \dodoi{10.1093/mnras/staf228}

\bibitem[{{Su} {et~al.}(2023){Su}, {Bryan}, {Haiman}, {Somerville}, {Hayward},
  \& {Faucher-Gigu{\`e}re}}]{Su+2023}
{Su}, K.-Y., {Bryan}, G.~L., {Haiman}, Z., {et~al.} 2023, \mnras, 520, 4258,
  \dodoi{10.1093/mnras/stad252}

\bibitem[{{Su} {et~al.}(2025{\natexlab{b}}){Su}, {Natarajan}, {Cho}, {Narayan},
  {Hopkins}, {Angl{\'e}s-Alc{\'a}zar}, \& {Prather}}]{Su+2025Bridging}
{Su}, K.-Y., {Natarajan}, P., {Cho}, H., {et~al.} 2025{\natexlab{b}}, \apjl,
  981, L33, \dodoi{10.3847/2041-8213/adb7dd}

\bibitem[{{Su} {et~al.}(2019){Su}, {Hopkins}, {Hayward}, {Ma},
  {Faucher-Gigu{\`e}re}, {Kere{\v{s}}}, {Orr}, {Chan}, \& {Robles}}]{Su+2019}
{Su}, K.-Y., {Hopkins}, P.~F., {Hayward}, C.~C., {et~al.} 2019, \mnras, 487,
  4393, \dodoi{10.1093/mnras/stz1494}

\bibitem[{{Su} {et~al.}(2021){Su}, {Hopkins}, {Bryan}, {Somerville}, {Hayward},
  {Angl{\'e}s-Alc{\'a}zar}, {Faucher-Gigu{\`e}re}, {Wellons}, {Stern},
  {Terrazas}, {Chan}, {Orr}, {Hummels}, {Feldmann}, \& {Kere{\v{s}}}}]{Su+2021}
{Su}, K.-Y., {Hopkins}, P.~F., {Bryan}, G.~L., {et~al.} 2021, \mnras, 507, 175,
  \dodoi{10.1093/mnras/stab2021}

\bibitem[{{Su} {et~al.}(2024){Su}, {Bryan}, {Hayward}, {Somerville}, {Hopkins},
  {Emami}, {Faucher-Gigu{\`e}re}, {Quataert}, {Ponnada}, {Fielding}, \&
  {Kere{\v{s}}}}]{Su+JetHaloMass2024}
{Su}, K.-Y., {Bryan}, G.~L., {Hayward}, C.~C., {et~al.} 2024, \mnras, 532,
  2724, \dodoi{10.1093/mnras/stae1629}

\bibitem[{{Talbot} {et~al.}(2021){Talbot}, {Bourne}, \& {Sijacki}}]{Talbot2021}
{Talbot}, R.~Y., {Bourne}, M.~A., \& {Sijacki}, D. 2021, \mnras, 504, 3619,
  \dodoi{10.1093/mnras/stab804}

\bibitem[{{Tamura} {et~al.}(2001){Tamura}, {Kaastra}, {Peterson}, {Paerels},
  {Mittaz}, {Trudolyubov}, {Stewart}, {Fabian}, {Mushotzky}, {Lumb}, \&
  {Ikebe}}]{Tamura+2001}
{Tamura}, T., {Kaastra}, J.~S., {Peterson}, J.~R., {et~al.} 2001, \aap, 365,
  L87, \dodoi{10.1051/0004-6361:20000038}

\bibitem[{{Tchekhovskoy} {et~al.}(2011){Tchekhovskoy}, {Narayan}, \&
  {McKinney}}]{Tchekhovskoy2011}
{Tchekhovskoy}, A., {Narayan}, R., \& {McKinney}, J.~C. 2011, \mnras, 418, L79,
  \dodoi{10.1111/j.1745-3933.2011.01147.x}

\bibitem[{{Torrey} {et~al.}(2020){Torrey}, {Hopkins}, {Faucher-Gigu{\`e}re},
  {Angl{\'e}s-Alc{\'a}zar}, {Quataert}, {Ma}, {Feldmann}, {Keres}, \&
  {Murray}}]{Torrey+2020}
{Torrey}, P., {Hopkins}, P.~F., {Faucher-Gigu{\`e}re}, C.-A., {et~al.} 2020,
  \mnras, 497, 5292, \dodoi{10.1093/mnras/staa2222}

\bibitem[{{Voit} {et~al.}(2015){Voit}, {Donahue}, {Bryan}, \&
  {McDonald}}]{Voit+2015}
{Voit}, G.~M., {Donahue}, M., {Bryan}, G.~L., \& {McDonald}, M. 2015, \nat,
  519, 203, \dodoi{10.1038/nature14167}

\bibitem[{{Weinberger} {et~al.}(2018){Weinberger}, {Springel}, {Pakmor},
  {Nelson}, {Genel}, {Pillepich}, {Vogelsberger}, {Marinacci}, {Naiman},
  {Torrey}, \& {Hernquist}}]{Weinberger2018}
{Weinberger}, R., {Springel}, V., {Pakmor}, R., {et~al.} 2018, \mnras, 479,
  4056, \dodoi{10.1093/mnras/sty1733}

\bibitem[{{Weinberger} {et~al.}(2023{\natexlab{a}}){Weinberger}, {Su},
  {Ehlert}, {Pfrommer}, {Hernquist}, {Bryan}, {Springel}, {Li}, {Burkhart},
  {Choi}, \& {Faucher-Gigu{\`e}re}}]{Weinberger2023}
{Weinberger}, R., {Su}, K.-Y., {Ehlert}, K., {et~al.} 2023{\natexlab{a}},
  \mnras, 523, 1104, \dodoi{10.1093/mnras/stad1396}

\bibitem[{{Weinberger} {et~al.}(2023{\natexlab{b}}){Weinberger}, {Su},
  {Ehlert}, {Pfrommer}, {Hernquist}, {Bryan}, {Springel}, {Li}, {Burkhart},
  {Choi}, \& {Faucher-Gigu{\`e}re}}]{Weinberger+2023}
---. 2023{\natexlab{b}}, \mnras, 523, 1104, \dodoi{10.1093/mnras/stad1396}

\bibitem[{{Wellons} {et~al.}(2023){Wellons}, {Faucher-Gigu{\`e}re}, {Hopkins},
  {Quataert}, {Angl{\'e}s-Alc{\'a}zar}, {Feldmann}, {Hayward}, {Kere{\v{s}}},
  {Su}, \& {Wetzel}}]{Wellons2023}
{Wellons}, S., {Faucher-Gigu{\`e}re}, C.-A., {Hopkins}, P.~F., {et~al.} 2023,
  \mnras, 520, 5394, \dodoi{10.1093/mnras/stad511}

\bibitem[{{Werner} {et~al.}(2013){Werner}, {Oonk}, {Canning}, {Allen},
  {Simionescu}, {Kos}, {van Weeren}, {Edge}, {Fabian}, {von der Linden},
  {Nulsen}, {Reynolds}, \& {Ruszkowski}}]{Werner+2013}
{Werner}, N., {Oonk}, J.~B.~R., {Canning}, R.~E.~A., {et~al.} 2013, \apj, 767,
  153, \dodoi{10.1088/0004-637X/767/2/153}

\bibitem[{{Wetzel} {et~al.}(2012){Wetzel}, {Tinker}, \& {Conroy}}]{Wetzel+2012}
{Wetzel}, A.~R., {Tinker}, J.~L., \& {Conroy}, C. 2012, \mnras, 424, 232,
  \dodoi{10.1111/j.1365-2966.2012.21188.x}

\end{thebibliography}
\bibliographystyle{aasjournal}


\end{CJK}
\end{document}